\documentstyle[emulateapj]{article}

\def\wig#1{\mathrel{\hbox{\hbox to 0pt{%
          \lower.5ex\hbox{$\sim$}\hss}\raise.4ex\hbox{$#1$}}}}

\def\Dwa{$\,$\uppercase\expandafter{\romannumeral5}$\,$}

\def\sles{\lower2pt\hbox{$\buildrel {\scriptstyle <}
   \over {\scriptstyle\sim}$}}
\def\sgreat{\lower2pt\hbox{$\buildrel {\scriptstyle >}
   \over {\scriptstyle\sim}$}}
\def\sharpnull#1{}

\slugcomment{Accepted to the Ap.J.}

\begin{document}

\title{Albedo and Reflection Spectra of Extrasolar Giant Planets}

\author{David Sudarsky\altaffilmark{1}, Adam Burrows\altaffilmark{1}, \&  Philip Pinto\altaffilmark{1}}

\altaffiltext{1}{Department of Astronomy and Steward Observatory, 
                 The University of Arizona, Tucson, AZ \ 85721}

\begin{abstract}

We generate theoretical albedo and reflection spectra
for a full range
of extrasolar giant planet (EGP) models, from Jovian to
51-Pegasi class objects.  Our albedo modeling utilizes the latest atomic
and molecular cross sections, Mie theory treatment of scattering and
absorption by condensates, a variety of particle size distributions,
and an extension of the Feautrier technique which allows for a general
treatment of the scattering phase function.

We find that due to qualitative similarities in the compositions and spectra of
objects within each of five
broad effective temperature ranges, it is natural to establish five
representative EGP albedo
classes.  At low effective temperatures
(T$_{\textrm{eff}} \lesssim$ 150 K) is
a class of ``Jovian'' objects (Class I) with tropospheric ammonia clouds. 
Somewhat warmer Class II, or ``water cloud,'' EGPs 
are primarily affected by condensed H$_2$O.
Gaseous methane absorption features are prevalent in both classes.
In the absence of non-equilibrium condensates in the upper atmosphere,
and with sufficient H$_2$O
condensation, Class II objects are expected to have the highest visible
albedos of any class.

When the upper atmosphere of an EGP is too hot for H$_2$O to condense, radiation
generally penetrates more deeply. 
In these objects, designated Class III or ``clear'' due to a lack of condensation
in the upper atmosphere,
absorption lines of the alkali metals,
sodium and potassium, lower the
albedo significantly throughout the visible.  Furthermore, the
near-infrared albedo is negligible, primarily due to strong CH$_4$ and H$_2$O
molecular absorption, and collision-induced
absorption (CIA) by H$_2$ molecules.
In those EGPs with exceedingly small orbital
distance (``roasters'') and 900 K $\lesssim$ T$_{\textrm{eff}} \lesssim$ 1500 K (Class IV), a tropospheric silicate layer is expected to exist.
In all but the hottest (T$_{\textrm{eff}} \gtrsim$ 1500 K) or lowest gravity
roasters, the effect of this silicate layer is
insignificant due to the very strong absorption
by sodium and potassium atoms above the layer.  The
resonance lines of sodium and potassium are expected to be salient features
in the reflection spectra of these EGPs.
In the absence of non-equilibrium condensates, 
we find, in contrast to previous studies, that these
Class IV roasters likely have the lowest visible and Bond albedos of any class, rivaling the
lowest albedos of our solar system.  For the small fraction of
roasters with T$_{\textrm{eff}} \gtrsim$ 1500 K and/or low surface gravity
($\lesssim 10^3$ cm s$^{-2}$; Class V), the
silicate layer is located very high in the atmosphere, reflecting much of
the incident radiation before it can reach the absorbing alkali metals and
molecular species.  Hence, the Class V roasters have
much higher albedos than those of Class IV.

We derive Bond albedos ($A_B$) and T$_{\textrm{eff}}$
estimates for the full set of known EGPs.  A broad range in both values is
found, with T$_{\textrm{eff}}$ ranging from $\sim$ 150 K to nearly 1600 K, and
$A_B$ from $\sim$ 0.02 to 0.8.

We find that variations in particle size distributions and
condensation fraction can have large quantitative, or even
qualitative, effects on albedo spectra.  In general, less condensation,
larger particle sizes, and wider size distributions result in lower
albedos.  We explore
the effects of non-equilibrium condensed products of photolysis
above or within principal cloud decks.  As in Jupiter, such species can lower
the UV/blue albedo substantially, even if present in relatively
small mixing ratios.

\end{abstract}

\section{Introduction}

Since the discovery of the extrasolar giant planet (EGP), 51 Pegasi b,
in 1995 (Mayor \& Queloz 1995), an explosion of similar discoveries
has followed.  To date, there are
$\sim$ 30 known planets orbiting nearby stars, which have collectively
initiated the new field of extrasolar giant planet research.

While to date most detections have been via Doppler
spectroscopy, other promising methods, both ground-based and
space-based, are in development.  These
include (but are not limited to) astrometric techniques (Horner et al.
1998), nulling
interferometry (Hinz et al. 1998), coronographic imaging (Nakajima
1994), and spectral deconvolution (Charbonneau et al. 1998).
Furthermore, planned space instrumentation such as the NGST (Next Generation
Space Telescope) and SIM (Space Interferometry Mission) may prove to be useful
for the detection and characterization of EGP systems.

With the current push for new instruments and techniques, we expect
that some of these new EGPs will soon be directly detected.  One
group (Cameron et al. 1999) has claimed a detection in reflected
light of the ``roaster,''
$\tau$ Boo b, while another group (Charbonneau et al. 1999)
has not claimed a detection, but has quoted an upper limit to the
albedo which
is in conflict with Cameron et al.
Our theoretical models of
EGP albedos are motivated by and can help guide attempts to
directly detect EGPs in reflection by
identifying their characteristic spectral features and by
illuminating the systematics.

The theoretical study of EGP albedos and reflection spectra is
still largely in its infancy.
Marley et al. (1999) have explored a range of stratosphere-free EGP geometric
and Bond albedos using temperature-pressure profiles of EGPs in isolation
(i.e. no stellar insolation), while
Goukenleuque et al. (1999) modeled 51 Peg in radiative equilibrium,
and Seager \& Sasselov (1998) explored radiative-convective models
of EGPs under strong stellar insolation.  In the present study,
our purpose is to provide
a broader set of models than previous work, and to establish a general understanding
of the albedo and reflection spectra of EGPs over the full
range of effective temperatures (T$_{\textrm{eff}}$).
Rather than attempting to model these spectra
in a fully consistent way for the
almost endless combinations of EGP masses, ages, orbital distances, elemental
abundances, and stellar spectral types, we concentrate on representative
composition classes based loosely on
T$_{\textrm{eff}}$.  
The ``Jovian'' Class I objects (T$_{\textrm{eff}} \lesssim$ 150 K)
are characterized by the presence of ammonia clouds.
(Note that the term, ``Jovian'', is used
here for convenience, not to imply that this entire class of objects will
be identical to Jupiter.)  In somewhat warmer objects (T$_{\textrm{eff}}
\sim$ 250 K), ammonia is in its
gaseous state, but the upper troposphere contains condensed H$_2$O.
These objects are designated Class II, or ``water cloud'' EGPs.
Class III, or ``clear'' EGPs, are so named because they are
too hot (T$_{\textrm{eff}} \gtrsim$ 350 K)
for significant H$_2$O condensation and so are not expected to contain
any principal condensates, though they are not necessarily completely cloud-free.
The hotter EGPs (900 K $\lesssim$ T$_{\textrm{eff}}$ $\lesssim$ 1500 K; Class IV) include those
objects with very small orbital distances (``roasters'') or those at large distances which are
massive and young enough to have similar effective temperatures.  In either case,
the troposphere of such an EGP is expected to contain significant abundances
of neutral sodium and potassium gases, as well as a silicate cloud layer.  The hottest
(T$_{\textrm{eff}} \gtrsim$ 1500) and/or least massive (g $\lesssim 10^3$ cm s$^{-2}$)
have a silicate layer located so high in the atmosphere that much of
the incoming radiation is shielded from alkali metal and molecular
absorption.

We use a planar asymmetric Feautrier method
in conjunction with
temperature-pressure (T-P) profiles, equilibrium gas abundances
(assuming Anders \& Grevesse (1989) elemental abundances),
and simple cloud models to account for condensed species.
The T-P profiles of isolated EGPs, as well as profiles
which are nearly isothermal in the outer atmosphere, are utilized.  This
allows us to bracket the effects of various T-P profiles
on the resulting EGP albedo spectra.  Like Marley et al. (1999),
we generate model albedo and reflection spectra and Bond albedos,
assuming a variety of central star spectral types.  Similarly, Rayleigh scattering,
Raman scattering, Mie extinction due to condensates, and
molecular absorption by a host of species are treated.  In addition to
our broader range of compositions and T$_{\textrm{eff}}$ than in Marley et al., we
treat the important absorption effects of the alkali metals, include a larger number of relevant condensates (including some
non-equilibrium products typical of photolysis), and produce a synthetic
albedo spectrum of Jupiter which is in reasonable agreement with Jupiter's actual albedo spectrum (Karkoschka 1994) from the soft UV to the near infrared.
 
Doppler spectroscopy favors the detection of massive companions at
small orbital distances and indeed EGPs with very small orbital radii have
been found.  $\tau$ Boo b (Butler et al. 1997), 51 Peg b
(Mayor \& Queloz 1995), $\upsilon$ And b (Butler et al. 1997),
HD 75289b (Mayor et al. 1999),
HD 187123b (Butler et al. 1998), HD 217107b (Fischer et al. 1999), and HD 209458b
(Charbonneau et al. 1999) all have orbital
distances of less than 0.1 AU and
masses (actually M$_p\sin i$) ranging from $\sim$0.4 to 3.4 Jupiter
masses.
Under stellar insolation, the elevated temperatures of EGPs depend mostly on the
level of stellar insolation, rather than on their masses and ages, which would
largely determine their T$_{\textrm{eff}}$ in isolation.  Using simple radiative
equilibrium arguments (T$_{\textrm{eff}} \propto F_{inc}^{1/4}$, where $F_{inc}$ is
the incident stellar flux),
most of the EGPs within 0.1 AU are likely to have very high T$_{\textrm{eff}}$
($\sim$ 800 K to over 1600 K). T$_{\textrm{eff}}$ is only weakly dependent
on the Bond albedo for a large range of low-to-moderate albedos, varying only
$\sim$ 20\% as the Bond albedo varies from 0.01 to 0.6.

At the other end of the scale, several objects with more traditional orbital
distances of $\gtrsim$ 1 AU have been discovered.  These EGPs include
16 Cyg Bb (Cochran et al. 1997), 47 UMa b (Butler et al. 1996),
$\upsilon$ And d (Marcy et al. 1999), Gl 614b (Mayor et al. 1998),
HR 5568b,
HR 810b, and HD 210277b (Marcy et al. 1998), and have M$_p\sin i$ ranging
from $\sim$ 0.75 to 5 M$_J$.
At these larger orbital distances, EGPs
receive much less stellar radiation and, therefore, have a
lower T$_{\textrm{eff}}$
($\lesssim$ 300K).  Still, many other EGPs,
such as 70 Vir b (Butler \& Marcy 1996), Gl 86 Ab (Queloz et al. 1999), and
HD 114762b (Latham et al 1989), have  
orbital distances between 0.1 and 1 AU and M$_p\sin i$
between 0.7 and 10 M$_J$.
Over the full set of currently known EGPs,
spectral classes of the central stars range from F7V to M4V.

The albedo of an object is simply the fraction of light that the object reflects.
However, there are several different types of albedos.  The {\it geometric\/} albedo
refers to the reflectivity of the object at full phase ($\Phi = 0$, where $\Phi$
represents the object's phase angle) relative to that
by a perfect Lambert disk of the same radius under the same incident flux.  Since planets are essentially
spheres, the factor projecting a unit surface onto a disk orthogonal to the line of
sight is given by $\cos\phi\sin\theta$, where $\phi$ is the object's longitude (defined
to be in the observer-planet-star plane) and $\theta$ is its polar angle
($\pi\over 2$ - latitude).  The
geometric albedo is given by integrating over the illuminated hemisphere:
\begin{equation}
A_g = {1\over {\pi I_{inc}}}\int_{\phi = {-{\pi\over 2}}}^{\pi\over 2}
\int_{\theta = 0}^\pi I(\phi,\theta,\Phi=0)\cos\phi\sin\theta d\Omega\, ,
\label{geometric}
\end{equation}
where $I_{inc}$ is the incident specific intensity, $\pi I_{inc}$ is the
incident flux, and $I(\phi,\theta,\Phi = 0)$ is the emergent intensity.
More generally, $I = I(\phi,\theta,\Phi;\phi_0,\theta_0)$, but at full phase all
incident angles ($\phi_0$, $\theta_0$) are equal to the emergent ones.  The
geometric albedo is usually given as a function of wavelength, although
it is sometimes averaged over a wavelength interval and stated as a single
number.

The {\it spherical\/} albedo, $A_s$,
refers to the fraction of incident light reflected by a sphere at all angles.  Usually stated as a function
of wavelength, it is obtained by integrating the reflected flux over all phase angles.
The flux ($F(\Phi)$) as a function of phase angle ($\Phi$) is given
by the more general form of eq.
(\ref{geometric}).  Assuming unit radius (Chamberlain \& Hunten 1987),
\begin{equation}
F(\Phi) = \int_{\phi = \Phi-{\pi\over 2}}^{\pi\over 2}\int_{\theta = 0}^\pi
I(\phi,\theta,\Phi;\phi_0,\theta_0)\cos\phi\sin\theta d\Omega.
\end{equation}

The spherical albedo is obtained by integrating $F(\Phi)$ over all
solid angles:
\begin{equation}
A_s = {1\over {\pi I_{inc}}}\int_{4\pi} F(\Phi) d\Omega =
{2\over I_{inc}} \int_0^\pi F(\Phi)\sin\Phi d\Phi.
\end{equation}

Note that the spherical and geometric albedos are related by $A_s = A_gq$, where
\begin{equation}
q = {2\over {F(\Phi=0)}}\int_0^\pi F(\Phi)\sin\Phi d\Phi
\end{equation}
is known as the phase integral.

The {\it Bond\/} albedo, $A_B$, is the ratio of the total reflected and total incident
powers.  It is obtained by weighting the spherical albedo by the spectrum of the
illuminating source and integrating over all wavelengths:
\begin{equation}
A_B = {\int_0^\infty A_{s,\lambda} I_{inc,\lambda} d\lambda
\over {\int_0^\infty I_{inc,\lambda} d\lambda}}\, ,
\label{bondeq}
\end{equation}
where the $\lambda$ subscript signifies that the incident intensity varies
with wavelength.

Spherical, geometric, and Bond albedos of objects are strong functions of their compositions.
Within the solar system, they vary substantially with wavelength,
and from object to object.  At short wavelengths, gaseous atmospheres can
have high albedos due to
Rayleigh scattering, and low albedos at longer wavelengths
due to molecular ro-vibrational absorption.  Icy condensates, whether they reside
on a surface or are present in an upper atmosphere, are highly reflective
and increase the albedo.  Other condensates, such as silicates or non-equilibrium
products of photolysis, can lower the albedo substantially over a broad wavelength region.

Some of the lowest albedos seen in the Solar System are those of asteroids
containing large amounts of carbonaceous material.  Many have Bond albedos of
less than
0.03 (Lebofsky et al 1989).
The Bond albedo of the Earth is
0.30 (Stephens et al. 1981) and that of the Moon is 0.11 (Buratti 1996). 
Jupiter and Saturn have somewhat higher Bond albedos, both near 0.35
(Conrath et al. 1989).

In \S \ref{modelsec}, we describe our approach to
modeling EGPs.  Section \ref{transfersec} describes our radiative transfer method,
\S \ref{moleculesec} contains a discussion of molecular absorption and scattering,
and \S \ref{condensatesec} describes
the properties of and our treatment of the relevant condensates in EGP
atmospheres.  In \S \ref{Jupitersec}, we discuss the application of
our methods to Jupiter,
\S \ref{resultssec} contains our EGP model albedo and reflection spectra
results, as well as T$_{\textrm{eff}}$ and Bond albedo estimates for
currently known EGPs,  and \S \ref{parametersec} describes
the effects of varying key parameters of the models.  We summarize our
results in \S \ref{conclusionsec}.

\section{Extrasolar Giant Planet Models \label{modelsec}}

Depending upon their proximity to their central stars as well as their
masses and ages, EGP effective temperatures likely span a large
range, from below 100 K to well over 1600 K, with highly varying
temperature-pressure-composition profiles.  However, an EGP's
outer atmospheric composition, rather than its specific
temperature-pressure profile, is
of primary importance in modeling albedos and reflection spectra.
With our composition classes, we encompass the range of behaviors
of EGP albedos and reflection spectra.  We do not model emission spectra,
nor do our models account for object-specific details, such as
elemental abundance differences or cloud patchiness.
EGPs are surely at least as rich and varied as the planets of our solar system,
but simple modeling reveals many interesting systematics.

\subsection{Temperature-Pressure Profiles}

Ideally, temperature (T)-pressure (P) profiles are computed directly via
radiative  equilibrium models of EGPs under
stellar insolation.  A move toward such models for very strong stellar insolation
has been made by
Seager \& Sasselov (1998) and Goukenleuque et al. (1999), while
for lower temperature objects,
Marley et al. (1999) utilize T-P profiles of
isolated EGPs.  The main effect of stellar insolation on the T-P profile
of an EGP is to make the outer atmosphere more nearly isothermal.
Studies of strong stellar insolation conclude that a stratosphere
does not exist in the high-temperature roasters (Seager \& Sasselov 1998;
Goukenleuque et al. 1999).  However,
it is not completely clear what might occur in the upper atmosphere if
ultraviolet photochemical processes are fully modeled.  Under solar insolation,
Jupiter and Saturn do exhibit stratospheres, and we suspect that the Class I 
EGPs are likely to have stratospheres as well.  In an albedo spectrum, the
existence of a
stratosphere is made manifest mainly by the scattering
and absorption effects of non-equilibrium aerosols which reside there.
Additionally, photochemical processes in the stratosphere may be the origin
of ``chromophores,'' non-equilibrium solids 
which settle near or in the upper cloud layers and are largely
responsible for the coloration of Jupiter.  

To bracket the range of albedos under stellar insolation, we use two sets
of pressure-temperature profiles.  The first
is a subset of profiles for theoretical isolated objects
(Marley et al. 1999; Marley 1998; Burrows et al. 1997) with T$_{\textrm{eff}} \approx$ 130 K (representing
an isolated Class I EGP),
250 K (Class II), 600 K (Class III), and 1200 K (Class IV).
We estimate that these representative isolated T-P profiles are valid
for surface gravities
between $\sim 3 \times 10^3$ to $3 \times 10^4$ cm s$^{-2}$.  A set of
modified profiles is obtained by altering these isolated
profiles to simulate a stellar insolated T-P profile by using the
model results of Seager \& Sasselov (1998) as a guide.  To approximate
the T-P profiles of the very hottest close-in objects (Class V),
we scale the 1200 K profile up to 1700 K.
We stress that these modified
profiles are very approximate, but along with the isolated T-P profiles, they
bracket a broad range of possible EGP T-P profiles. 

Figure \ref{tpprofiles} shows both the isolated and modified T-P profiles for
Classes I through IV, as well as our modified Class V profile.  Also shown are condensation curves, which indicate the highest
temperatures and pressures at which species can condense.  Cloud bases are
located approximately where the profiles intersect
the condensation curves (dotted curves). 
Class I (``Jovian'') objects contain both ammonia and deeper water cloud
layers, while water is likely the only principal condensate present in the
tropospheres of Class II objects. (As shown in Figure \ref{tpprofiles},
a thin ammonia haze layer might appear very high in the atmosphere for
an isolated Class II T-P profile.)   The Class III 
T-P profile doesn't cross any principal condensation curves in the
upper atmosphere, regardless of the
level of stellar insolation.  Finally, the Class IV and V roasters
contain a silicate cloud deck and a deeper iron cloud deck throughout
the full range of possible T-P profiles, though their
cloud depths differ considerably.  

\subsection{Determination of Gaseous Abundances \label{gasabunsec}}

Using the analytic formulae in Burrows \& Sharp (1999), we calculate
gaseous mixing ratios of the main compounds of carbon, oxygen,
and nitrogen (CH$_4$, CO, H$_2$O,
NH$_3$, N$_2$) over the full range of temperatures
and pressures in the model EGP atmospheres.
H$_2$ and He mixing ratios are set according to Anders \& Grevesse (1989) solar
abundances, and the
H$_2$S mixing ratio is set in accordance with the Anders \& Grevesse
abundance of sulfur ($\sim 3 \times 10^{-5}$).  The abundances of the
alkali metals (Na, K, Rb, Cs), important in the Class III through Class V
EGPs, are calculated numerically
using the formalism of Burrows \& Sharp (1999).  

Overall, the effect of differences in the T-P profile on gaseous mixing ratios tends to
be greatest for the Class IV objects due to the temperature and pressure
dependences of neutral alkali metal mixing ratios and the fact that
the T-P profiles are in the
vicinity of the CH$_4$/CO and NH$_3$/N$_2$ equilibrium curves.
From
the standpoint of gaseous abundances, the T-P profiles have little effect on
the albedos of cooler EGPs.

\subsection{Cloud Modeling \label{cloudsec}}

Our treatment of clouds in our fiducial EGP models assumes that the gaseous form
of a condensable species is completely depleted above the cloud deck and
that the species settles within the cloud layer in its condensed form.  The base
of the cloud resides where the T-P profile of the EGP
meets the condensation curve of the given species, and the cloud top
is simply set at one pressure scale height above the base.  Not all of the
given condensable within the cloud is in condensed form.  Rather, at the
base of the cloud, the gaseous form is assumed to be at the saturation
vapor pressure.
For a given abundance of a condensable, if we assume that the portion
of the condensable which exceeds the 
saturation vapor pressure is entirely in condensed form, we will
refer to this as ``full condensation.''  However, as in Jupiter's
outer atmosphere (see \S
\ref{Jupitersec}), it
is possible that the condensation fraction will be smaller.
Hence, we retain the condensation fraction 
as a parameter.  Furthermore, the particle size distributions in EGP
atmospheres are impossible to ascertain at this point, so particle
size remains a free parameter as well.

The standard model for Jupiter lends some support to our prescription
for clouds.  The
base of Jupiter's ammonia cloud deck resides approximately
where its T-P profile
meets the NH$_3$ condensation curve ($\sim$ 0.7 bar) and the cloud tops
extend roughly one pressure scale height, to $\sim$ 0.3 bar
(West et al. 1986; Griffith et al. 1992).  Although
present, NH$_3$ gas is largely depleted above
the cloud layer.

In the case of silicate condensation, where the condensate and gas molecules
are not identical (unlike NH$_3$ and H$_2$O), the condensate abundance is estimated
by the Anders \& Grevesse abundance of the limiting
species.  We use enstatite (MgSiO$_3$; though a number of other silicates with differing
optical properties are certainly present), for which the limiting element is silicon.  For the
full condensation limit, it is assumed
that the entire mass of silicon above the pressure of the cloud base
settles into MgSiO$_3$ within the cloud.

\section{Radiative Transfer Method \label{transfersec}}

Due to the forward scattering from condensates in EGP atmospheres, an appropriate
radiative transfer method must allow for a forward-backward asymmetric scattering
phase function.  Although the conventional Feautrier method (e.g. Mihalas 1978) does not
allow for such an asymmetry, a straightforward extension of this technique can be derived by separating the
source function into upward- and downward-propagating rays (Mihalas 1980; Milkey et al. 1975).

At first thought, it may seem inappropriate to use a planar
transfer code in the modeling of albedos and reflection spectra from
spherical objects.  However, it is fairly straightforward to derive the equivalence
between uniform radiation from one direction onto a unit sphere and
uniform radiation from 2$\pi$ steradians onto a plane with unit area.
Hence, provided that we set the incident intensity to be uniform in angle, the
spherical albedo is the ratio of the outward and incident fluxes.

The fundamental transfer equation is
\begin{equation}
\mu {\partial I(\mu) \over \partial\tau} = I(\mu) - S(\mu),
\end{equation}
where the source function is given by
\begin{equation}
S(\mu) = {1\over 2}\sigma\int_{-1}^{1}R(\mu,\mu^\prime)I(\mu^\prime)d\mu^\prime
\end{equation}
and the thermal term is neglected in this albedo study.  $R(\mu,\mu^\prime)$ is the
 azimuth-independent angular redistribution function (azimuthal symmetry
is assumed) and $\sigma$ is the single-scattering
albedo, $\sigma = \sigma_{scat}/\sigma_{ext}$, where $\sigma_{scat}$ is the
scattering cross section and $\sigma_{ext}$ is the extinction cross section.
Separated into upward ($I^+$) and downward ($I^-$)
components, the transfer equation becomes
\begin{equation}
\mu{\partial I^+(\mu)\over\partial\tau} = I^+(\mu) - S^+(\mu)
\label{tplus}
\end{equation}
and
\begin{equation}
-\mu{\partial I^-(\mu)\over\partial\tau} = I^-(\mu) - S^-(\mu),
\label{tminus}
\end{equation}
where the source functions are given by
\begin{equation}
S^+(\mu) = {1\over 2}\sigma\int_0^1 \left[R(\mu,\mu^\prime) I^+(\mu^\prime) +
R(\mu,-\mu^\prime)I^-(\mu^\prime)\right]d\mu^\prime
\end{equation}
and
\begin{equation}
S^-(\mu) = {1\over 2}\sigma\int_0^1 \left[R(-\mu,\mu^\prime) I^+(\mu^\prime) +
R(-\mu,-\mu^\prime)I^-(\mu^\prime)\right]d\mu^\prime
\end{equation}
for the $I^+$ and $I^-$ equations, respectively.

Forming symmetric and antisymmetric averages, and using the Feautrier
variables, $u = {1\over 2}(I^+ + I^-)$ and $v = {1\over 2}(I^+ - I^-)$, eqs.
(\ref{tplus}) and (\ref{tminus}) are rewritten as
\begin{equation}
\mu{\partial v(\mu)\over\partial\tau} = u(\mu) - {1\over 2}\left[S^+(\mu)
+ S^-(\mu)\right]
\label{mudvdt}
\end{equation}
and
\begin{equation}
\mu{\partial u(\mu)\over\partial\tau} = v(\mu) - {1\over 2}\left[S^+(\mu)
- S^-(\mu)\right].
\label{mududt}
\end{equation}

Since $R(\mu,\mu^\prime)$ depends only upon the angle between $\mu$ and
$\mu^\prime$, the following symmetries exist:
\begin{eqnarray}
R(\mu,\mu^\prime) = R(-\mu,-\mu^\prime) \\
R(-\mu,\mu^\prime) = R(\mu,-\mu^\prime).
\end{eqnarray}

With the definitions, $R^+(\mu,\mu^\prime) = R(\mu,\mu^\prime) + 
R(-\mu,\mu^\prime)$ and $R^-(\mu,\mu^\prime) = R(\mu,\mu^\prime) -
R(-\mu,\mu^\prime)$, eqs. (\ref{mudvdt}) and (\ref{mududt}) become
\begin{equation}
\mu{\partial v\over\partial\tau} = u - {1\over 2}\sigma\int_0^1 R^+
(\mu,\mu^\prime)u(\mu^\prime)d\mu^\prime
\end{equation}
and
\begin{equation}
\mu{\partial u\over\partial\tau} = v - {1\over 2}\sigma\int_0^1 R^-(\mu,\mu^\prime)
v(\mu^\prime)d\mu^\prime.
\end{equation}

This system of first-order equations is discretized for numerical
computation by replacing the derivatives with difference quotients, and
by substituting Gaussian quadrature sums for the integrals.  The principal
equations then become
\begin{equation}
\mu_i{{v_{d,i}-v_{d-1,i}}\over \Delta\tau_d} = u_{d,i} - {1\over 2}\sigma
\sum_j\omega_j R^+(\mu_i,\mu_j) u_{d,j}
\label{disc1}
\end{equation}
and
\begin{equation}
\mu_i{{u_{d+1,i}-u_{d,i}}\over \Delta\tau_{d+{1\over 2}}} = v_{d,i}
-{1\over 2}\sigma\sum_j\omega_j R^-(\mu_i,\mu_j) v_{d,j}\, ,
\label{disc2}
\end{equation}
where $d$ signifies a given depth zone ($d = 1,...,D$), and $i$ and $j$ signify angular bins ($i,j = 1,...,N$).
$\Delta\tau_{d+{1\over 2}}$ equals $\tau_{d+1} - \tau_d$ and $\Delta\tau_d$ equals ${1\over 2} 
(\Delta\tau_{d+{1\over 2}} + \Delta\tau_{d-{1\over 2}})$.  To achieve
numerical stability, $\Delta\tau$ is
staggered by half a zone in eq. (\ref{disc2}) relative to eq. (\ref{disc1}).  
The $\omega_j$ are the Gaussian weights.

The upper boundary conditions are given by the relations, $u_{1,i} - v_{1,i} =
I^-_i$ and
\begin{equation}
\mu_i{{u_{2,i}-u_{1,i}}\over\Delta\tau_{1\over 2}} = u_{1,i} - I^-_i
- {1\over 2}\sigma\sum_j\omega_j \left[u_{1,j}-I^-_j\right] R^-(\mu_i,\mu_j)\, ,
\end{equation}
where $I^-_i$  and $I^-_j$ signify the incident intensity as a function of
angle at the surface.  We set $I^-$ to unity at all angles since only the
{\it ratio\/} of the outward and inward fluxes determines the spherical
albedo.  The lower
boundary conditions are given by $u_{D,i} + v_{D,i} = I^+_i$ and
\begin{equation}
\mu_i{{u_{D,i}-u_{D-1,i}}\over\Delta\tau_{D-{1\over 2}}} = I^+_i - u_{D,i}
- {1\over 2}\sigma\sum_j\omega_j \left[I^+_j - u_{D,j}\right] R^-(\mu_i,\mu_j)\, ,
\end{equation}
where $I^+_i$ and $I^+_j$ signify the outward-traveling intensity at the base of the atmosphere
(set to zero in this study).

The system of equations can be represented by angle
matrices (${\bf A}_d, {\bf B}_d, {\bf C}_d,...$) and column vectors (${\bf u}_d$ and
${\bf v}_d$) such that equations (\ref{disc1}) and (\ref{disc2}) can be written as
\begin{equation}
{\bf A}_d{\bf v}_{d-1} + {\bf B}_d{\bf u}_d + {\bf C}_d{\bf v}_d = 0
\end{equation}
and
\begin{equation}
{\bf D}_d{\bf u}_d + {\bf E}_d{\bf v}_d + {\bf F}_d{\bf v}_{d+1} = 0.
\end{equation}

Given D depth zones and N angles, the system results in a block matrix
containing $[2\times D]^2$ submatrices, each of order N.  Implementing the
boundary conditions described above, this system is solved directly via
LU decomposition and substitution.

Our atmosphere models utilize 100 optical depth zones with logarithmic
sizing near the surface, where higher resolution is essential, and a continuous
transition to linear zoning at depth.  Sixteen polar angular bins per hemisphere
are used.

\subsection{Quantitative Comparison for Uniform Atmospheres}

In order to test our asymmetric Feautrier code, we compare our resulting
spherical albedos for uniform atmosphere models with those derived employing both
Monte Carlo and analytic techniques.  Van de Hulst (1974) derived
a solution for the spherical albedo of a planet covered with a semi-infinite
homogeneous cloud layer.  Given a single-scattering albedo of $\sigma$
(= $\sigma_{scat}/\sigma_{ext}$) and a scattering asymmetry factor of $g
= <\cos\theta>$ (the average cosine of the scattering angle), van de
Hulst's expression for the spherical albedo is
\begin{equation}
A_s \approx {(1 - 0.139s)(1 - s)\over {1 + 1.170s}}\, ,
\end{equation}
where
\begin{equation}
s = \left[{1 - \sigma}\over {1 - \sigma g}\right]^{1/2}.
\end{equation}

Figure \ref{vancompare} shows the spherical albedo
of a homogeneous, semi-infinite atmosphere as a function
of scattering asymmetry factor and single scattering albedo.  Along with
van de Hulst's semi-analytic curves are our asymmetric Feautrier
and Monte
Carlo model results using
a Henyey-Greenstein scattering phase function,
\begin{equation}
p(\theta) = {{1-g^2}\over {(1+g^2-2g\cos\theta)^{3/2}}}.
\end{equation}

For nearly all values of $g$ and $\sigma$, the agreement is very good,
differing by under 1\%.  There
are slightly larger variations when both $g$ and $\sigma$ approach unity due to
the finite number of angles and depth zones used in our numerical models,
but in actual planetary or EGP atmospheres, this corner of parameter space is rarely realized.

Real planetary atmospheres are usually highly stratified and the optical
depth is a strong function of wavelength.  Given the atmospheric
 temperature-pressure-composition profile, an appropriate conversion to optical depth is
required.  Assuming hydrostatic equilibrium and using an ideal gas
equation of state, this conversion is
\begin{equation}
d\tau = {\sigma_{ext}(P)\over{\textrm{g}\mu(P)}}dP,
\end{equation}
where $\sigma_{ext}$ is the effective extinction cross section per
particle at depth $P$, $\mu$ is the mean molecular weight, and g is the surface
gravity (assumed constant because the depth of the effective atmosphere is a very
small fraction of the planet's radius).

\section{Atomic and Molecular Scattering and Absorption \label{moleculesec}}

The gases present in EGP atmospheres are many (Burrows and Sharp 1999).
However, only some of them have the requisite abundances and cross sections at
the temperatures and pressures of upper EGP atmospheres to have significant
spectral effects in the visible and near-infrared.  These species include
H$_2$, CH$_4$, H$_2$O, NH$_3$, CO, and H$_2$S.  Additionally, Na and K
are important absorbers in Class III, IV, and V EGPs. 

Of course, H$_2$ is the most abundant species, followed by helium.
The dominant carbon-bearing molecule is a function of both temperature and
pressure.  Chemical equilibrium modeling (Burrows and Sharp 1999;
Fegley \& Lodders 1996)
shows that, at solar metallicity,  CH$_4$ will dominate over CO in most EGP atmospheres.
At high temperatures, the CO abundance overtakes that of CH$_4$
($\sim$ 1100 K at 1 bar; $\sim$ 1400 K at 10 bars).  There is a similar transition
for the nitrogen-bearing molecules:  NH$_3$ dominates at low temperatures,
but it is overtaken by N$_2$ at higher temperatures ($\sim$ 700 K at 1 bar;
$\sim$ 900 K at 10 bars).
Some species condense into solids at low temperatures, thereby depleting the
gaseous phase.  In particular, NH$_3$ condenses below 150--200 K (depending upon pressure),
as does H$_2$O below 250--300 K. 

At visible and near-infrared wavelengths, molecular absorption is due to
ro-vibrational transitions, so molecular opacity is a very strong function
of wavelength.  Even when no permanent dipole moment exists, such as with
the H$_2$ molecule, the high gas pressures in EGP atmospheres can induce temporary
dipole moments via collisions.  This Collision Induced Absorption (CIA)
is responsible for broad H$_2$-H$_2$ (and H$_2$-He) absorption bands in
Jupiter and Saturn (Zheng \& Borysow 1995; Trafton 1967).

The temperature- and pressure-dependent gaseous opacities are obtained
from a variety of sources---a combination of
theoretical and experimental data as referenced in Burrows
et al. (1997).  Additionally, for this study the CH$_4$ opacity was extended
continuously into the visible wavelength region using the
data of Strong et al. (1993) and a methane absorption spectrum from Karkoschka (1994).
These two data sets were then extrapolated in temperature and pressure by
scaling with existing temperature and pressure-dependent near-infrared CH$_4$
data (Burrows et al. 1997 and references therein).

Many prominent molecular absorption features may be seen in EGP albedo
and reflection spectra. 
At relatively low temperatures, broad H$_2$-H$_2$ and H$_2$-He CIA bands
peak at $\sim$ 0.8 $\mu$m, 
1.2 $\mu$m, and 2.4 $\mu$m.  At higher temperatures and pressures, the CIA
cross sections become larger at
all wavelengths.  CIA is especially
important in cloud-free gaseous objects, where incident radiation is absorbed
deeper in the atmosphere.  NH$_3$ absorption bands shortward of 2.5 $\mu$m
occur at $\sim$ 1.5 $\mu$m,
2.0 $\mu$m, and 2.3 $\mu$m.  (Note that our database does not contain the visible
bands of ammonia.)  H$_2$O absorption occurs at $\sim$ 0.6 $\mu$m, 0.65 $\mu$m, 0.7 $\mu$m,
0.73 $\mu$m, 0.82 $\mu$m, 0.91 $\mu$m, 0.94 $\mu$m, 1.13 $\mu$m, 1.4 $\mu$m,
1.86 $\mu$m, and 2.6 $\mu$m.  A large number of CH$_4$
features appear in the visible and near-infrared.  Some of the more
prominent ones occur at $\sim$ 0.54 $\mu$m, 0.62 $\mu$m, 0.67 $\mu$m, 0.7 $\mu$m, 0.73 $\mu$m, 0.79 $\mu$m, 0.84 $\mu$m,
0.86 $\mu$m, 0.89 $\mu$m, 0.99 $\mu$m, 1.15 $\mu$m, 1.4 $\mu$m, 1.7 $\mu$m, and 2.3 $\mu$m.  CO absorption bands occur
at $\sim$ 1.2 $\mu$m, 1.6 $\mu$m, and 2.3 $\mu$m and 
H$_2$S features may be found at $\sim$ 0.55 $\mu$m, 0.58 $\mu$m, 0.63 $\mu$m, 0.67 $\mu$m, 0.73 $\mu$m, 0.88 $\mu$m,
1.12 $\mu$m, 1.6 $\mu$m, and 1.95 $\mu$m.  Of course, depending upon
mixing ratios and cross sections, only some of these features will appear in a given
EGP albedo spectrum.

Strong pressure-broadened lines of neutral sodium and potassium are expected
to dominate the visible albedos of Class III and Class IV EGPs. 
The most prominent absorption lines
of sodium occur at 3303 \AA, 5890 \AA, and 5896 \AA,
while those of potassium occur at 4044 \AA, 7665 \AA, and 7699 \AA.

Atomic and molecular scattering includes conservative Rayleigh scattering as
well as non-conservative Raman scattering.  In the case of Rayleigh
scattering, cross sections are derived from polarizabilities,
which are in turn derived from refractive indices.
Since the refractive indices are readily available at 5893 \AA\ 
(Weast 1983), the Rayleigh cross sections are derived at this wavelength
via,
\begin{equation}
\sigma_{Ray} = {8\over 3}\pi k^4 \left({{n-1}\over{2\pi L_0}}
\right)^2\, ,
\end{equation}
where $k$ is the wavenumber at this wavelength
(2$\pi/\lambda \simeq 106621$ cm$^{-1}$) and $L_0$ is Loschmidt's number.
Assuming that the refractive indices are not strong functions of
wavelength, we simply extrapolate these cross sections as $\lambda^{-4}$.
  
Raman scattering by H$_2$ involves the shift of continuum photons to longer
or shorter wavelengths as they scatter off H$_2$, exciting or de-exciting
rotational and vibrational
transitions.  Raman scattering is not coherent in frequency, so a rigorous
treatment is not possible with our transfer code.  Instead, we adopt the
approximate method introduced by Pollack et al. (1986) and used by Marley
et al. (1999) in their albedo study.
At a given wavelength, the single scattering albedo within
a particular depth zone is approximated by
\begin{equation}
\sigma = {{\sigma_{Ray}+\sigma^{\prime}_{scat}+
(f_{{\lambda}^*}/f_{\lambda})\sigma_{Ram}}\over {\sigma_{Ray}+\sigma^{\prime}_{ext}
+\sigma_{Ram}}},
\end{equation}
where $f_\lambda$ denotes the spectrum of incident radiation (the spectrum of an
EGP's central star), $\lambda^{* -1} = \lambda^{-1} + \Delta
\lambda^{-1}$, where $\Delta\lambda$ is the wavelength of the H$_2$
vibrational fundamental ($\Delta\lambda^{-1}$ = 4161 cm$^{-1}$),
$\sigma_{Ram}$ is the Raman cross section,
and $\sigma^{\prime}_{scat}$ and $\sigma^{\prime}_{ext}$ are the
effective condensate scattering and extinction cross sections, respectively.
Raman scattering may be significant in deep gaseous planetary atmospheres, where it
can lower the UV/blue albedo by up to $\sim$ 15\% (Cochran \& Trafton 1978).
However, our models show that in higher temperature EGP atmospheres,
alkali metal absorption can dominate
over this wavelength region, while in cooler EGP atmospheres, condensates
largely dominate.  Over our full set of EGP models, we find that Raman
scattering is relatively insignificant.

\section{Mie Theory and Optical Properties of Condensates \label{condensatesec}}

Condensed species in EGP atmospheres range from ammonia ice in low temperature objects
to silicate grains at high temperatures.  Some of the condensates relevant
to EGP atmospheres include NH$_3$ ($\lesssim$ 150--200K), NH$_4$SH
($\lesssim$ 200K), H$_2$O ($\lesssim$ 250--300K), low-abundance
sulfides and chlorides ($\lesssim$ 700--1100K),
silicates such as MgSiO$_3$ ($\lesssim$ 1600--1800K), and iron or iron-rich compounds
($\lesssim$ 1900--2300K).  Additionally, photochemical processes in
the upper atmosphere can produce non-equilibrium condensates.
Stratospheric hazes may be composed of
polyacetylene (Bar-Nun et al. 1988) and other aerosols.  Chromophores, those non-equilibrium
species which cause the coloration of Jupiter and Saturn, might include P$_4$ (Noy et al. 1981) or
organic species similar to Titan tholin (Khare \& Sagan 1984).

Condensates can have drastic effects on EGP reflection spectra, increasing the albedo
at most wavelengths, but sometimes depressing the albedo in the UV/blue.  Of course,
those condensates which are higher in the atmosphere will
generally have a greater effect than those which reside more deeply.  The presence
and location of a particular condensed species is determined largely by an object's
T-P profile, and by the tendency of the condensate to settle
(due to rain) at a depth in the atmosphere near the region where the T-P profile crosses the condensation curve.
Hence, a given low-temperature (T$_{\textrm{eff}} \lesssim$ 150K) atmosphere
might consist of an ammonia cloud deck high in the troposphere and a water cloud deck
somewhat deeper, with purely gaseous regions above, beneath, and between the clouds.
Similarly, a high-temperature (T$_{\textrm{eff}} \sim$ 1200K) atmosphere might consist of a 
tropospheric silicate cloud deck above a deeper iron cloud deck.  Depending upon the
amount of condensate in the upper cloud and the wavelength region, the presence of deeper clouds may or may not have
any effect on the albedo and reflection spectrum.

The scattering and absorption of electromagnetic radiation by condensed species in planetary
atmospheres is a very complex problem.  The extinction properties of ices, grains, and droplets of
various sizes, shapes, and compositions cannot be described accurately by simple
means.  Most often, these properties are approximated by Mie Theory, which describes the solution of Maxwell's
equations inside and outside a homogeneous sphere with a given complex refractive index.

We use a full Mie Theory approach which utilizes the formalism of
van de Hulst (1957) and Deirmendjian (1969), and results in scattering
and extinction cross sections as well as a scattering asymmetry
factor, $g = <\cos\theta>$, given the complex index of refraction and
particle radius ($a$).  Larger particles require an increasing
number of terms in an infinite series to describe these parameters
accurately, and so they require more computing time.  But
while the cross sections and scattering asymmetry factors of
small- to moderately-sized particles ($2\pi a/\lambda \lesssim 75$) vary substantially with wavelength, these
variations are greatly reduced for larger spheres.  For these larger
particles, we use an asymptotic form of the Mie equations outlined
fully by Irvine (1965). Interpolation between the full Mie
theory results and these asymptotic limits yields the parameters for large
particles.  However, inherent assumptions in the asymptotic form of the Mie equations
render them inadequate for the computation of the scattering cross sections in the
weak-absorption limit ($n_{imag} \lesssim  10^{-3}$), in which case we use
the geometric optics approximation (Bohren \& Huffman 1983),
\begin{equation}
Q_{sca} = 2 - {8\over 3}{n_{imag}\over n_{real}}\left[n_{real}^3 - \left(
n_{real}^2 - 1 \right)^{3/2}\right]x \ ,
\end{equation}
where $Q_{sca}$ is the usual scattering coefficient (the ratio of the
scattering cross section to the geometric cross section), $x$ is the size
parameter ($=2\pi a/\lambda$), $n_{real}$ is the real index of refraction,
and $n_{imag}$ is the imaginary component of the refractive index.

The principal condensates to which we have applied Mie theory include NH$_3$ ice,
H$_2$O ice, and MgSiO$_3$ (enstatite), where the
optical properties, namely the complex indices of refraction, were obtained from
Martonchik et al. (1984), Warren (1984), and Dorschner et al. (1995), respectively.  The
complex refractive indices of NH$_3$ were interpolated in the 0.7 to 1.4 $\mu$m
wavelength region, due to the lack of data there.

Each of these condensates
has absorption features, as is made evident by the behavior
of the imaginary index of refraction (Figure \ref{indices1}).  Shortward of 2.5 $\mu$m,
NH$_3$ ice absorption occurs
at $\sim$ 1.55 $\mu$m, 1.65 $\mu$m, 2.0 $\mu$m, and 2.25 $\mu$m. H$_2$O ice produces broader
features at $\sim$ 1.5 $\mu$m and 2.0 $\mu$m.  Enstatite is mostly featureless
below 2.5 $\mu$m, except shortward of $\sim$ 0.35 $\mu$m.



The non-equilibrium condensates to which we have applied Mie theory include  
phosphorus (Noy et al. 1981), tholin (Khare \& Sagan 1984),
and polyacetylene (Bar-Nun et al. 1988).  P$_4$ and tholin are
chromophore candidates, particularly for the coloration of Jupiter and Saturn,
due to their large imaginary indices of refraction in the UV/blue
(Figure \ref{indices2}) and
plausibility of production.
A somewhat yellowish allotrope of phosphorus, P$_4$ was produced in the laboratory by Noy et al.
(1981) by UV irradiation of an H$_2$/PH$_3$ gaseous mixture.  It is believed
that this same process may be responsible for its production in Jupiter.
Tholin is a dark-reddish organic solid
(composed of over 75 compounds) synthesized by Khare and Sagan (1984) by
irradiation of gases in a 
simulated Titan atmosphere.  It is believed that a tholin-like solid may be
produced similarly in giant planet atmospheres.  Polyacetylenes, polymers of
C$_2$H$_2$, were investigated by Bar-Nun et al. (1988) and likely are an optically
dominant species in the photochemical stratospheric hazes of giant planets,
where hydrocarbons are abundant (Edgington et al. 1996; Noll et al. 1986).  

Cloud particle sizes are not easily modeled and are a strong function of
the unknown meteorology in EGP atmospheres.  Inferred particle sizes in
solar system giant planet atmospheres can guide EGP models, though they range
widely from fractions of a micron to tens of microns.

We have investigated various particle size distributions.  A commonly used
distribution, and the one that we use in our fiducial models, is
\begin{equation}
n(a) \propto \left({a\over a_0}\right)^6 \exp\left[-6\left({a\over a_0}\right)
\right],
\label{cloud}
\end{equation}
which reproduces the distributions in cumulus water clouds in Earth's
atmosphere fairly well if the
peak of the distribution is
a$_0 \sim 4 \mu$m (Deirmendjian 1964).
Stratospheric aerosols---at least those in Earth's stratosphere---can be represented by the ``haze'' distribution
(Deirmendjian 1964),
\begin{equation}
n(a) \propto {a\over a_0}\exp\left[-2\left({a\over a_0}\right)^{1/2}\right].
\end{equation}

\section{The Albedo of Jupiter \label{Jupitersec}}

Jupiter is an important testbed for the theory of albedos, since
full-disk geometric albedo spectra have been obtained (Karkoschka 1994,
1998), and because space-based and ground-based studies have provided
a fair amount of information concerning Jupiter's atmosphere.    
At visible and near-infrared wavelengths, Jupiter's
upper troposphere and stratosphere shape
its albedo spectrum.
According to the standard model, a somewhat heterogeneous cloud deck extends
from $\sim$ 0.3 to 0.7 bars in the troposphere (West et al. 1986;
Griffith et al. 1992).  Although the bulk of the cloud deck consists
primarily of particles at least 10 $\mu$m in size, a layer
of smaller particles ($\sim$ 0.5--1.0 $\mu$m) resides near the cloud
tops (West et al. 1986; Pope et al. 1992).  Beneath this upper cloud
deck is a NH$_4$SH and NH$_3$ cloud layer at $\sim$ 2--4 bars and an
H$_2$O cloud condenses somewhat deeper.  Above the NH$_3$ cloud deck, a
stratospheric haze resides at pressures near $\sim$ 0.1 bar.  It is
worth mentioning that the Galileo probe results deviate
from this standard model.  One difference relevant to the albedo
and reflection spectra is a
tropospheric haze inferred from the probe data, likely composed primarily of NH$_3$,
above a somewhat deeper NH$_3$
cloud deck (Banfield et al. 1998),
but it is not known whether the
probe entry location is characteristic of the planet as a whole.

In addition to H$_2$, abundant gaseous species in the upper troposphere include
He and CH$_4$, with mixing ratios relative to H$_2$ of
0.156 and $\sim 2.1 \times 10^{-3}$, respectively (Niemann et al. 1996).
Gaseous NH$_3$, H$_2$O, H$_2$S, and PH$_3$ are present in small mixing ratios.

It is suggested that the color differences of Jupiter's belts and
zones are largely due to the visibility of chromophores residing within
the NH$_3$ cloud deck (West et al. 1986).  No appreciable altitude differences
between the belts and zones are found, although the zones likely
contain thicker upper cloud and/or haze layers than the belts (Chanover
et al. 1997; Smith 1986).  Jupiter's UV/blue albedo is depressed
substantially from what one would expect from the increase with frequency of the Rayleigh
scattering cross sections, and Raman scattering cannot account for the
albedo in this wavelength region.  This depressed UV/blue albedo likely is
produced by the
large imaginary refractive indices of the tropospheric chromophores and, to a lesser
degree, stratospheric aerosols (West et al. 1986).

Due to the large optical depth of Jupiter's upper ammonia cloud deck at
visible and near-infrared wavelengths, a
two-cloud model of the atmosphere suffices (West 1979; Kuehn \& Beebe 1993).
We model the top of the upper cloud deck ($\sim$ 0.35 bar) with a ``cloud''
distribution (see \S \ref{condensatesec}) peaked at 0.5 $\mu$m.  Deeper in the cloud, from
0.45 to 0.7 bar, a particle distribution peaked at 30 $\mu$m is used.  This size
distribution is also utilized in the lower cloud, spanning 2 to 4 bars.

In addition
to NH$_3$ condensation, a small mixing ratio of a chromophore, either tholin
($2 \times 10^{-8}$) or P$_4$ ($5 \times 10^{-9}$), is added to the
upper cloud.  As inferred from
limb darkening observations, the condensed chromophore becomes well mixed
in the upper ammonia cloud, and perhaps deeper as well (West et al. 1986;
Pope et al. 1992).  As per Noy et al. (1981), the peak of the chromophore
particle size distribution is set to 0.05 $\mu$m.  However, the nature
of the size distribution and whether the chromophore adheres to
the ammonia ice particles are as yet unclear. 

Gaseous abundances are modeled using the Galileo Probe Mass Spectrometer values
(Niemann et al. 1996) as a guide.  The H$_2$, He, and CH$_4$ abundances are taken
directly from the Probe results.
However, the tropospheric NH$_3$ abundance varies
considerably with depth.  At $\sim$ 0.4 bar, its mixing ratio has been found
to be $\sim 5 \times 10^{-6}$ (Griffith et al. 1992; Kunde et al. 1982),
while at
$\sim$ 0.7 bar, its mixing ratio is $\sim 5 \times 10^{-5}$ to $10^{-4}$.
In an infrared study using Voyager IRIS data (Gierasch et al. 1986), it was
found that only a small fraction ($\sim$ 1\%) of the ammonia is in condensed
form.  Based on our visible albedo modeling, where the smaller particle size
distribution dominates, and using the gaseous NH$_3$
mixing ratios above, we find that a condensation fraction of $\sim$ 5\%
in the upper cloud is required to provide the necessary reflectance.

We model Jupiter's stratospheric haze using Deirmendjian's ``haze''
particle size distribution (see \S \ref{condensatesec}) of polyacetylene
peaked at 0.1 $\mu$m---a
particle size justified by limb darkening studies (Rages et al. 1997;
West 1988; Tomasko et al. 1986).  The abundance of C$_2$H$_2$ in Jupiter's
stratosphere is $\sim 10^{-8}$ to $10^{-7}$ (Edgington 1998; Noll et. al. 1986),
though the polymerized abundance is not known.  In this study, the
polyacetylene mixing ratio is set to $5 \times 10^{-8}$, in a haze
layer from 0.03 to 0.1 bar.

Figure \ref{Jupiter} shows two model geometric albedo spectra along with the
observed full-disk albedo spectrum of Jupiter (Karkoschka 1994).  We convert our
model spherical albedo to a geometric albedo using an averaged phase integral
of $q$ = 1.25 (Hanel et al. 1981).  The upper
model utilizes tholin as the chromophore throughout the upper ammonia
cloud deck, while the lower
model utilizes P$_4$.

Although the general character of Jupiter's geometric albedo is reproduced
fairly well, many of the methane absorption features are modeled too
deeply.  Furthermore, the gaseous ammonia
features at $\sim$ 0.65 $\mu$m and 0.79 $\mu$m do not appear in the models because
our database does not include ammonia absorption shortward of $\sim$
1.4 $\mu$m.  Karkoschka (1998) indicates that the absorption
feature centered at $\sim$ 0.93 $\mu$m may be due to ammonia as well.
Relying upon Mie scattering theory and our choices for chromophore
particle size distributions, tholin appears to reproduce the UV/blue
region of the albedo better than P$_4$.  However, the actual
chromophore(s) in Jupiter's atmosphere remains a mystery.  

The published Bond albedo of Jupiter is 0.343 (Hanel et al. 1981).
Using our models and limited wavelength coverage (0.3 $\mu$m to 2.5 $\mu$m), we
estimate a Bond albedo (see \S \ref{resultssec}) in the 0.42 to 0.44
range---a fair approximation---depending upon whether P$_4$ or tholin is used as the
chromophore.

Uncertainties in the vertical structure of Jupiter's atmosphere,
heterogeneities in Jupiter's cloud layers, and our use of an averaged
phase integral all likely play a role in explaining the
differences between observational and modeled albedo spectra.
These details aside, Jupiter's
atmosphere remains a useful benchmark for our models of EGP albedo and
reflection spectra.

\section{Results for EGPs \label{resultssec}}

We produce fiducial albedo models for each EGP class using both isolated
and modified temperature-pressure profiles.  We adopt Deirmendjian's ``cloud''
particle size distribution with a peak
at the moderate size of 5 $\mu$m.  Our model EGP spherical albedos for
the full range of effective temperatures
are shown in Figures
\ref{spherical1} through \ref{spherical3}.  For these fiducial models,
``full condensation'' is assumed (as described in
\S \ref{cloudsec}).

The ``Jovian'' Class I albedo spectra are determined mainly by the
reflectivity of condensed NH$_3$ and the molecular absorption bands of
gaseous CH$_4$.  Stratospheric and tropospheric non-equilibrium species
are not included
in these fiducial models.  Their effects are explored in \S
\ref{parametersec}.  Because both isolated and nearly ``isothermal''
T-P profiles of EGPs with T$_{\textrm{eff}} \lesssim$ 150 K cross the NH$_3$
condensation curve, the details of the T-P profiles do not have a
large impact on the resulting albedos of Class I objects.
As shown in Figure \ref{spherical1}a, the reflective
NH$_3$ clouds keep the albedo fairly
high throughout most of the visible spectral region.  Toward the infrared,
the gaseous absorption cross sections tend to become larger,
so photons are more likely to be absorbed above the cloud deck.  Hence,
at most infrared wavelengths, the albedo is below that in the visible region.
      
The isolated and modified profile Class II albedos are shown in Figure \ref{spherical1}b.
Relative to a Class I EGP, a Class II albedo is even higher in the visible due to
very strongly reflective H$_2$O clouds in the upper atmosphere.  Gaseous
absorption features tend to be shallower because these H$_2$O clouds
form higher in the atmosphere than the NH$_3$ clouds of most Class I objects.
The intersection of the isolated profile and the NH$_3$ condensation curve
near 0.01 bars may result in a thin NH$_3$ condensation layer high in the
atmosphere, but NH$_3$ condensation is assumed to be negligible for this model.

The ``clear'' Class III does not contain any principal condensates
in the upper atmosphere (irrespective of the T-P profile), although
a silicate cloud deck exists deeper, at $\sim$ 50 bars.
The presence of alkali metals in the troposphere has 
a substantial lowering effect on the albedo.  As per Figure
\ref{spherical2}a, sodium and potassium absorption
lowers the albedo at short wavelengths, resulting in
a spherical albedo below $\sim$ 0.6 throughout most of the UV/blue
spectral region.  Into the red region,
lower Rayleigh scattering cross sections and strong alkali metal
absorption result in spherical albedos which drop below 0.1.
In contrast, in the absence of the alkali metals, the
spherical albedo would remain high ($\gtrsim$ 0.75) throughout most of the
visible. In both cases, the
near-infrared albedo is essentially negligible, largely due to absorption by
CH$_4$, H$_2$O, and H$_2$-H$_2$ CIA.  Our models show that, in Class III objects, the details of
the T-P profile will have only minor effects on the albedo.
If low-abundance sulfide or chloride condensates were to exist in the troposphere,
they could appear at pressures as low as a few bars.  Based on
theoretical abundances (Burrows \& Sharp 1999), thick clouds are very
unlikely, but it is worth mentioning that even cirrus-like condensation
could raise the albedo in the visible and near-infrared.

In the higher-temperature (900 K $\lesssim$ T$_{\textrm{eff}} \lesssim$ 1500 K)
Class IV roasters, the effect of the alkali
metals is most dramatic.
Unlike the Class III EGPs,
a silicate cloud deck exists
at moderate pressures of $\sim$ 5--10 bars, depending on the details
of the T-P profile.  An iron or iron-rich condensate likely 
exists below the silicate deck, but it is sufficiently below
the opaque silicate cloud that it does not have any effect on the
visible and near-infrared albedos.  Figure \ref{spherical2}b shows
the spherical albedo of a Class IV EGP.  Assuming a fairly ``isothermal''
T-P profile (the modified profile) in the upper atmosphere, absorption by sodium and
potassium atoms, coupled with ro-vibrational molecular absorption,
results in a surprisingly low albedo
throughout virtually the entire visible and near-infrared wavelength region
explored in this study ($\leq 2.5 \mu$m).  The silicate
cloud is deep enough that its effects are rendered negligible by the
absorptive gases above it.  

Although a Class IV model with a modified T-P profile results in an albedo
which is significantly lower than that of even a Class III model, the
albedo of
a Class IV model with an isolated T-P profile is a different story:  Because the upper
atmosphere in such a model is significantly cooler than in the modified T-P profile
case, the equilibrium abundances of the alkali metals are lower.  Furthermore, the
silicate cloud deck is expected to be somewhat higher in the atmosphere (Figure \ref{tpprofiles})
and to have a non-negligible effect on the albedo in both the visible and
near-infrared regions (Figure \ref{spherical2}b).

Due to the low ionization potentials of sodium (5.139 eV) and
potassium (4.341 eV), it is likely that significant Na II
and K II layers exist in the outer atmospheres of Class IV EGPs (and
perhaps Class III EGPs).
Nevertheless, assuming a silicate cloud layer at $\sim$ 5-10 bars, simple ionization
equilibrium estimates indicate that these layers should not reach
the depths of the silicate layer in Class IV EGPs, and so substantial
column depths of Na I and K I should remain to absorb visible radiation.  The
full absorption and emission features of such ionization layers will
be explored in future EGP studies.

The very hot (T$_{\textrm{eff}} \gtrsim$ 1500 K) Class V roasters
have a silicate cloud layer which is located much higher in the atmosphere
relative to the Class IV roasters, so alkali metal and molecular
absorption is reduced.  Figure
\ref{spherical2}b illustrates the much higher albedo expected
of the Class V objects, assuming the silicate
layer is composed predominantly of enstatite grains.  If the sodium and
potassium ionization layers are substantial in these objects, then
the absorption due to their neutral lines will be reduced even further.
We also note that a roaster of particularly low mass (e.g. HD 209458b)
is expected to exhibit
a significantly larger radius than such an object in isolation
(Burrows et al. 2000).  For such a low surface
gravity ($\lesssim 10^3$ cm s$^{-2}$) object, the silicate
layer will form high in the atmosphere even for T$_{\textrm{eff}}$
$< 1500$ K.  Hence, the lower limit to T$_{\textrm{eff}}$
required to render a roaster a Class V EGP is reduced in the
case of low surface gravity.


Via spectral deconvolution,
Charbonneau et al. (1999) have constrained
the geometric albedo of the roaster, $\tau$ Boo b, to be below
0.3 at 0.48 $\mu$m.  This limit, which was obtained with an assumed phase
function and orbital inclination near 90 degrees, contrasts with the
findings of Cameron et al. (1999), who infer that the albedo
is high in this region.  Using our Class IV T-P profile model
(T$_{\textrm{eff}} \lesssim$ 1500 K), we
find that the geometric albedo at 0.48 $\mu$m is only 0.03.  However, if
in fact $\tau$ Boo b is a Class V EGP (T$_{\textrm{eff}} \gtrsim$ 1500 K),
we derive a geometric albedo of 0.39 at 0.48 $\mu$m, still smaller than
the assumed Cameron et al. value of 0.55, from which they derive a planetary radius
as high as 1.8 Jupiter radii.  The widely varying
albedos of Classes IV and V coupled with the fact that $\tau$ Boo b appears
to have an effective temperature near the transition region between
these classes indicates that the detailed modeling of this
EGP will be necessary in order to ascertain its nature.

It is instructive to examine the temperatures and pressures to which
incident radiation penetrates an EGP's atmosphere as a function
of wavelength.  For each class, Figures \ref{tauoneP}
and \ref{tauoneT} show the
pressures and temperatures, respectively, corresponding to one mean free path of an
incident photon.
In clear atmospheres, these temperatures and pressures
are very strong functions 
of wavelength, largely mirroring molecular absorption bands and/or
atomic absorption lines.  Conversely,
when thick
cloud layers are present, the wavelength dependence is much weaker, due
to the efficient extinction of radiation by a size distribution of
condensed particles.

Due to the azimuthal symmetry of our Feautrier technique, we do not
compute the phase integrals of EGPs.
In their absence, the characteristics of the
atmosphere at $\tau_{\lambda}$ $\sim$ 1
are useful
for the approximate conversion from spherical albedos to geometric albedos.
Using the asymmetry factor and single scattering albedo 
values, the phase integral, $q_{\lambda}$,
is estimated by interpolating
within the tables of Dlugach \& Yanovitskij (1974) (Marley et al. 1999).
Geometric albedos
are then obtained using the relation,
 $A_{g,\lambda} = A_{s,\lambda}/q_{\lambda}$.
Estimated geometric albedo spectra are shown in
Figure \ref{geoplot}. 
Our Class II 
geometric albedo compares qualitatively with that
of the ``quiescent'' water cloud model of Marley et al. (1999).  Given the
differences in particle size distributions, the Marley et al. albedo
tends to fall off a bit more sharply with increasing wavelength, while
having shallower gaseous absorption features in the visible.  Our
Class IV models may be compared with the Marley et al.
``brown dwarf'' model with silicate (enstatite) clouds, as well as with
the 51-Peg b model of Goukenleuque et al. (1999).
Our inclusion
of the alkali metals results in a qualitatively very different, and much lower, albedo
spectrum than in these previous studies.  Our Class V model may be compared
with the high-temperature model (T$_{\textrm{eff}}$ = 1580 K) of
Seager \& Sasselov (1998).  We find that, similar to Seager \& Sasselov,
the presence of silicate (enstatite) grains results in significant reflection,
but our inclusion of the alkali metals results in prominent absorption lines
as well. 

We combine a geometric albedo spectrum from each EGP class with appropriately
calibrated stellar
spectra (Silva \& Cornell 1992) to produce representative EGP reflection spectra.  Figure
\ref{refplot}a shows theoretical full-phase reflection spectra of EGPs from
0.35 $\mu$m to 1.0 $\mu$m, assuming
a G2V central star, orbital distances of 0.05 AU (Class IV), 0.2 AU (Class III),
1.0 AU (Class II), and 5.0 AU (Class I), and a planetary radius of 1 Jupiter radius
(R$_J$).
For a Class IV roaster, at 0.45 $\mu$m, the ratio of reflected and stellar
fluxes is
$\sim$ $5 \times 10^{-6}$, while for Class I, II, and III EGPs, it is $\sim$
$5 \times 10^{-9}$, $10^{-7}$, and $10^{-6}$, respectively.  This ratio
at 0.65 $\mu$m drops to $\sim$ $5 \times 10^{-7}$ for a Class IV
object, and is $\sim$ $5 \times 10^{-9}$, $10^{-7}$, and $3 \times 10^{-7}$ for
Class I, II, and III EGPs, respectively.  Figure  \ref{refplot}b shows theoretical
full-phase reflection spectra of Class IV and V roasters, assuming an F7V central star,
orbital distances of 0.1 AU (Class IV) and 0.04 AU (Class V), and 1 R$_J$.  At
0.45 $\mu$m, the reflected to stellar flux ratios are $\sim$ $10^{-6}$ (Class IV)
and $5 \times 10^{-5}$ (Class V).  At 0.65 $\mu$m, these ratios are
$\sim$ $10^{-7}$ (Class IV) and $5 \times 10^{-5}$ (Class V).  For larger planetary radii and
different orbital distances, these ratios should be scaled accordingly. 

In the reflection spectrum of a Class IV object (Figure \ref{refplot}),
absorption by the
resonance lines of sodium (5890\AA/5896\AA) and potassium
(7665\AA/7699\AA) is extreme.
These lines are
also very significant, though substantially weaker, in Class III objects.
Methane absorption bands shortward of 1 $\mu$m, especially those at $\sim$ 0.73 $\mu$m,
0.86 $\mu$m, and 0.89 $\mu$m, are quite prominent in Class I and III objects.
These bands are also clearly present in Class II objects, but with
sufficient water condensation high in the troposphere, the bands are not
as prominent as in Class I or Class III EGPs.  Although present, methane
absorption is even weaker in Class IV EGPs, where CO is the dominant carbon-bearing
molecule.  At the high effective temperature of a Class V object
(T$_{\textrm{eff}} \gtrsim$ 1500 K), the methane abundance is completely
overwhelmed by that of CO, and we expect that no strong methane bands will be seen in
reflection.

Bond albedos for EGPs are obtained using eq. (\ref{bondeq}).  Our lower and
upper wavelength limits of integration are 0.3 $\mu$m and 2.5 $\mu$m, respectively, rather
than formally from 0 to infinity.  Hence, our derivations are estimates
of actual Bond albedos, accurate to $\sim$ 10-15\%, depending on the
central stellar spectral type and the uncertainties in the EGP spherical
albedos shortward of 0.3 $\mu$m and longward of 2.5 $\mu$m.
The Bond albedos for
our fiducial modified T-P profile models and for isolated T-P profile
models are shown in Tables \ref{Bondtaba} and \ref{Bondtabb}.  Assuming full condensation of
principal condensates and no non-equilibrium species, the Bond albedos of
Class I and II objects are high.
Over the spectral range, M4V to A8V, the peak of the stellar energy
flux ranges from $\sim$ 0.9 $\mu$m to 0.4 $\mu$m.  Class I
EGP Bond albedos range from $\sim$ 0.4 to 0.65, while those of Class II EGPs
reach nearly 0.9.  These albedos tend to be significantly lower
when smaller condensation fractions and non-equilibrium condensates are
considered.  For example, the Bond albedo of our Jupiter model about
a G2V central star is in
the 0.42 to 0.44 range, depending upon whether P$_4$ or tholin is used
as the chromophore---somewhat higher than Jupiter's actual Bond
albedo of 0.343 (Hanel et al. 1981).

In contrast, Bond albedos of Class III and IV EGPs are very low.  Those of Class
III objects vary from $\sim$ 0.01 to 0.2 over the spectral range, M4V to A8V.
Class IV EGPs reflect the smallest fraction of
incident radiation, with
Bond albedos ranging from below
0.01 up to only 0.04, assuming our modified T-P profile model and no
non-equilibrium condensates.  These Bond albedos are significantly
lower than those of Marley et al. (1999) because we include the
effects of the alkali metals.  For example, assuming a G2V central
star, our fiducial Class III model yields a Bond albedo of 0.12,
while those of Marley et al. are in the 0.31 to 0.33 range
(cloud-free 500 K models), and our Bond albedo for a Class IV EGP
is only 0.03, while those of Marley et al. are in the 0.30 to
0.44 range (cloud-free and cloudy 1000K models).  The Bond albedos
of the very hot Class V objects are much higher than those
of Class III or IV, ranging from $\sim$ 0.51 to 0.57 over the the spectral
range, M4V to A8V.

Estimated Bond albedos and effective temperatures of known EGPs are
shown in Tables 2 through 4.  The equilibrium temperature of an
irradiated object is
\begin{equation}
T_{\textrm{eq}} = \left[{(1-A_B)L_{*}\over {16\pi\sigma a^{2}}}\right]^{1/4}
\end{equation}
(Saumon et al. 1996), where $L_{*}$ is the stellar luminosity, $\sigma$ is the Stefan-Boltzmann
constant, and $a$ is the orbital distance of the planet.  For massive
and young EGPs
with sufficiently large orbital distances, T$_{\textrm{eff}}$
$>$ T$_{\textrm{eq}}$ due to their significant internal energies.  We
estimate the effective temperatures of such objects simply by adding
the stellar-insolated and internal contributions to the luminosity, and
noting that $L = 4\pi R_p^2\sigma T_{\textrm{eff}}^4$.  The internal
contribution is defined to be the luminosity of an isolated object of
the given mass and age, and it is found using the evolutionary models of Burrows
et al. (1997).

Given the list of over two dozen known EGPs, it is possible that
none is cold enough to be a Class I (``Jovian'') object (HR 5568b is
an ambiguous case).  Classes II, III, and IV
are well-represented (Tables 2 through 4), while Class V likely includes
HD 209458b, and perhaps $\tau$ Boo b and/or HD 75289b.

\section{Parameter Studies \label{parametersec}}

In EGP atmospheres, variations in condensation fractions and particle
size distributions, as well as the possible presence of stratospheric
and tropospheric non-equilibrium species, can have large effects on the
spherical and Bond albedos.
First, we consider the effects of lowering the
condensation fraction to 10\% and 1\%.  Figures \ref{condfrac}a
and \ref{condfrac}b show
the substantial changes in Class I (``Jovian'') and Class II 
(``water cloud'') EGPs.  The Class II case best illustrates
the systematic effects, since only an H$_2$O cloud deck exists.  (Recall that the Class I model contains
an ammonia cloud deck above a water cloud deck.)
The condensation fraction has a
substantial effect on the spherical and geometric albedos.
Less condensation clearly results in lower albedos, especially
in the red/near-infrared, where gaseous opacities are strong (Marley
et al. 1999).  Note that the effects of the alkali metals, deep in the
atmosphere, are apparent in the UV/blue albedo of the Class II, 1\% condensation model.
As shown in Tables \ref{Bondtaba} and \ref{Bondtabb},
the Bond albedos of these 1\% condensation models are significantly lower
than those of their ``full condensation'' counterparts, particularly for 
the Class II EGPs. 

Cloud particle size distributions in EGPs are not known.  As alluded to
in \S \ref{condensatesec}, for a given condensate abundance, the net extinction by
condensates (almost pure scattering for H$_2$O ice)
is smaller when particle sizes are larger.  This is shown explicitly 
in Figure \ref{paramfig}a, comparing spherical albedos for
Deirmendjian H$_2$O ice ``cloud'' distributions
with size peaks of 0.5 $\mu$m, 5 $\mu$m (fiducial), and 50 $\mu$m.
The qualitative effect of increasing the peak size is
similar to the effect of reducing the condensation fraction.
Widening the distribution has similar consequences because the
largest particles squander the condensate, reducing the number
density of smaller scattering particles.

Non-equilibrium species in the upper atmospheres of EGPs may be produced
by UV-induced processes.  While both gaseous and condensed species
are likely to be produced, the condensates will generally
have greater effects on the albedos and reflection spectra.
As in the atmosphere of Jupiter, stratospheric hazes and tropospheric chromophores,
or impurities within or above the principal cloud layers, can lower the
albedo spectra in the UV/blue range and can also modify their
character at other wavelengths.  In addition to their compositions,
the size distributions of these non-equilibrium species play a role.
Figure \ref{paramfig}b shows the effect of including a representative
upper tropospheric ``haze'' of tholin (with mixing ratio of 10$^{-8}$) on
the spherical albedo of a Class I EGP.
In analogy with our Jupiter model,
the size distribution is peaked at 0.05 $\mu$m.  Although the abundances
and size distributions of such particles in EGPs are unknown,
we present this model as an indication of the qualitative effect
that this type of haze would have on the albedo.  The associated Class I
Bond albedo, assuming a G2V central star, decreases from 0.57 to 0.48.

We represent the optically dominant aerosol within stratospheric hazes by
polyacetylene, although other possibilities certainly exist.
Our models show that the effect of polyacetylene on the albedo is minor,
lowering the UV/blue albedo no more than a few percent, assuming a mixing
ratio as large as $10^{-7}$.
We stress that the actual compositions, abundances, and the size
distributions of non-equilibrium species in 
EGPs are unknown, and that
the quantitative effects on EGP albedos may or may not be significant.

\section{Conclusions \label{conclusionsec}}

The classification of EGPs into five composition classes,
related to T$_{\textrm{eff}}$, is instructive,
since the albedos of objects within each of these classes
exhibit similar features and values.  The principal condensate in Class I ``Jovian''
EGPs (T$_{\textrm{eff}} \lesssim$ 150 K) is NH$_3$, while in Class II
``water cloud'' EGPs it is
H$_2$O ice.  Gaseous molecular absorption features, especially
those of methane, are exhibited throughout
Class I and II albedo spectra.  Assuming adequate levels
of condensation, Class II EGPs are the most highly reflective of any class.
For lower condensation fractions, the albedos of both classes
fall off more quickly with increasing wavelength relative to ``full condensation''
models---especially the Class II objects. 
Even a small mixing ratio of a non-equilibrium tropospheric condensate within
or above a cloud deck can
depress the UV/blue albedo and reflection spectrum significantly.

In Class III ``clear'' EGPs (T$_{\textrm{eff}} \gtrsim$ 350 K), little condensation
is likely, and so albedos are determined almost entirely by atomic and molecular
absorption and Rayleigh scattering.  Radiation generally penetrates more deeply
into these atmospheres,
to pressures and temperatures where sodium and potassium absorption and H$_2$-H$_2$
collision-induced absorption (CIA) become substantial.  Throughout most of the
visible spectral region, the albedo decreases with increasing wavelength.   
In the near-infrared, CIA, H$_2$O, and CH$_4$ conspire to keep the albedo
very low.

In the upper atmospheres of the high-temperature (900 K $\lesssim$ T$_{\textrm{eff}} \lesssim$ 1500 K) Class IV roasters,
the equilibrium abundances of the alkali
metals are higher than in the Class III EGPs, so the absorption lines
of sodium and potassium are expected to lower the albedo more dramatically.
A silicate cloud exists at moderate depths ($\sim$ 5--10 bars), but the large
absorption cross sections of the sodium and potassium gases above it preclude
the cloud from having a significant effect on the albedo.  Like Class III EGPs,
the near-infrared albedo is expected to remain close to zero in the absence
of non-equilibrium condensates.

The hottest (T$_{\textrm{eff}} \gtrsim$ 1500 K) and/or lowest gravity
(g $\lesssim 10^3$ cm s$^{-2}$) roasters (Class V) have a silicate layer located much higher
in the atmosphere relative to the Class IV roasters.  This layer
is expected to reflect much of the incident radiation before it is
absorbed by neutral sodium and potassium and molecular species.  Hence, Class V EGPs
have much higher albedos than those of Class IV.


While stratospheres generally are not anticipated in high
temperature EGPs (Seager \&
Sasselov 1998; Goukenleuque et al. 1999), it is possible that
more detailed modeling will show that they do exist.  The
presence of a stratosphere would give rise to visible and
infrared emission features not otherwise seen.  Furthermore, the
presence of non-equilibrium solids due to photochemistry may decrease
the albedo in the UV/blue, but increase it somewhat in the red/near-infrared because even largely absorbing condensates are more
reflective than gaseous molecular species in this spectral region.

Differences in particle size distributions of the principal condensates
can have large quantitative, or even qualitative effects on the
resulting albedo spectra.  In general, less condensation, larger particle
sizes, and wider size distributions result in lower albedos.

Despite many uncertainties in the atmospheric details of EGPs, our set of
model albedo spectra serves as a useful guide to the prominent features and
systematics
over a full range of EGP effective temperatures, from $\sim$ 100 K to
1700 K.
Full radiative equilibrium modeling of a given EGP at a specific orbital distance
from its central star (of given spectral type), and of specific
mass, age, and composition is necessary for a detailed understanding of an
object. 
However, as observational EGP spectra become available, our set of model albedo
spectra offers
a means by which a quick understanding of their general character is
possible, and by which
some major atmospheric constituents,
both gaseous and condensed, may be inferred.

\acknowledgements
 
We thank Mark Marley, Sara Seager, Bill Hubbard, Jonathan Lunine,
Christopher Sharp,
and Don Huffman
for a variety of useful contributions.
This work was supported under NASA grants NAG5-7073 and NAG5-7499.

\clearpage

\figcaption{{
Temperature-pressure profiles for each of the EGP classes of this
study.  Both isolated and modified (more isothermal) profiles
are shown for Classes I through IV, as well as a modified profile
for Class V.
Also plotted are the condensation curves for some principal
condensates, as well as the NH$_3$/N$_2$ and CH$_4$/CO abundance
equilibrium curves.
}\label{tpprofiles}}

\figcaption{{
Comparison of our Asymmetric Feautrier code results to Monte Carlo and
analytic solutions for deep, homogeneous atmospheres.  The spherical albedo is plotted as a function of
the single-scattering albedo (= $\sigma_{scat}$/$\sigma_{ext}$) and
the average value of the cosine of the scattering angle ($g$ = $<\cos\theta>$).
}\label{vancompare}}

\figcaption{{
Imaginary refractive indices of the principal condensates used in this study.
}\label{indices1}}


  
\figcaption{{
Imaginary refractive indices of stratospheric haze and tropospheric chromophore
candidates.  Tholin and P$_4$ provide a great deal of absorption in
the UV/blue.
}\label{indices2}}

\figcaption{{
The geometric albedo spectrum of Jupiter.  Our model albedo spectra (thin curves)
are compared with the observational full-disk albedo spectrum (thick curve,
Karkoschka 1994).  The top model utilizes tholin as a chromophore,
while the bottom model uses P$_4$.
}\label{Jupiter}}

\figcaption{{
(a) The spherical albedo of a Class I ``Jovian'' EGP.  
Irrespective of the T-P profile, an NH$_3$ cloud deck resides above
an H$_2$O cloud deck.  The thin curve corresponds to an isolated T-P profile
model, while the thick curve signifies a modified, more isothermal
profile model. 
(b) The spherical albedo of a Class II ``water cloud'' EGP.
A thick H$_2$O cloud deck in the upper troposphere
produces a high albedo.
Isolated (thin curve) and modified (thick curve)
T-P profile models are shown.
}\label{spherical1}}

\figcaption{{
(a) The spherical albedo of a Class III ``clear'' EGP.
In addition to the isolated (thin curve) and modified
(thick curve) T-P profile models, the dashed curve depicts what the albedo
would look like in the absence of the alkali metals.
(b) The spherical albedo of a Class IV ``roaster.''
Theoretical albedo spectra of isolated (thin curve) and modified
(thick curve) T-P profile Class IV models are depicted.
}\label{spherical2}}

\figcaption{{
The spherical albedo of a Class V roaster.  A silicate layer high in
the atmosphere results in a much higher albedo than a Class IV roaster.
}\label{spherical3}}

\figcaption{{
The log of the pressure (P) at a depth equal to the mean free path
of incident radiation,
as a function of wavelength
($\lambda$) in microns, for
each of the EGP classes.  A size distribution of particles and the
assumption of ``full condensation'' conspire to make
the Class I and Class II curves only weak functions of wavelength.
}\label{tauoneP}}

\figcaption{{
The temperature (T) at a depth equal to the mean free path of incident radiation,
as a function of
wavelength ($\lambda$) in microns, for each of the EGP classes.
}\label{tauoneT}}

\figcaption{{
(a) Estimated geometric albedos of Class I, II and III EGPs.  A modified T-P profile model
is used in each case.  These conversions from spherical albedos are made
by approximating the phase integral ($q_{\lambda}$) based on the single
scattering albedo and scattering asymmetry factor at an atmospheric
depth equal to the mean free path of incident radiation.
(b) Estimated geometric albedos of Class IV and V EGPs.
}\label{geoplot}}

\figcaption{{
(a) Full-phase EGP reflection spectra, assuming a G2V central star, a planetary
radius equal to that of Jupiter, and orbital
distances of 0.05 AU (Class IV), 0.2 AU (Class III), 1.0 AU (Class II),
and 5.0 AU (Class I).
These reflection spectra are obtained by combining the geometric albedo
of each EGP class with a G2V stellar spectrum (Silva \&
Cornell 1992) and a Kurucz (1979) theoretical spectrum longward of 0.9 $\mu$m.
(b) Full-phase EGP reflection spectra, assuming an
F7V central star, a planetary radius equal to that of Jupiter, and orbital distances of 0.1 AU (Class IV) and 0.04 AU
(Class V).
}\label{refplot}} 

\figcaption{{
(a) The dependence of the spherical albedo of a Class I EGP on condensation
fraction.  ``Full condensation'' (thick curve), 10\% condensation
(thin curve), and 1\% condensation (dashed curve) models are shown.
(b) The dependence of the spherical albedo of a Class II EGP on condensation
fraction.  ``Full condensation'' (thick curve), 10\% condensation
(thin curve), and 1\% condensation (dashed curve) models are shown.
}\label{condfrac}}

\figcaption{{
(a) The dependence of the spherical albedo on the particle size distribution.
Class II EGP models with Deirmendjian ``cloud'' distributions peaked
at 0.5 $\mu$m (thin curve), 5$\mu$m (fiducial; thick curve), and
50 $\mu$m (dashed curve) are shown.  Note that the alkali metals are not
included in these model albedo spectra.
(b) The effect of the presence of an upper tropospheric tholin haze (with mixing
ratio of $10^{-8}$) on a Class I
EGP spherical albedo.  The albedo-lowering effect is greatest in the UV/blue
region of the spectrum.
}\label{paramfig}}

\clearpage

\begin{deluxetable}{cccccc}
\tablewidth{16cm}
\tablenum{1a}
\tablecaption{Estimated Bond Albedos of EGPs \label{Bondtaba}}
\tablehead{
\colhead{star} & \colhead{EGP class} & \colhead{fiducial}
& \colhead{``isolated''} & \colhead{10\% cond.}
& \colhead{1\% cond.}}
\startdata

A8V & I   & 0.63 & 0.64 & 0.62 & 0.59 \nl
    & II  & 0.88 & 0.88 & 0.79 & 0.47 \nl
    & III & 0.17 & 0.13 &     &     \nl
    & IV  & 0.04 & 0.21 &     &     \nl
    & V   & 0.57 &      &     &     \nl
F7V & I   & 0.59 & 0.61 & 0.57 & 0.51 \nl
    & II  & 0.84 & 0.83 & 0.74 & 0.40 \nl
    & III & 0.14 & 0.10 &     &     \nl
    & IV  & 0.03 & 0.18 &     &     \nl
    & V   & 0.56 &      &     &     \nl
G2V & I   & 0.57 & 0.59 & 0.55 & 0.47 \nl
    & II  & 0.81 & 0.81 & 0.71 & 0.37 \nl
    & III & 0.12 & 0.09 &     &     \nl
    & IV  & 0.03 & 0.16 &     &     \nl
    & V   & 0.55 &      &     &     \nl

\tablenotetext{}{Bond albedos of EGPs, using modified T-P profile models (fiducial)
with full condensation,
isolated T-P profile models (``isolated'') with full condensation,
modified T-P profile models with 10\% condensation (10\% cond.), and modified
T-P profile models with 1\% condensation (1\% cond.).  Because Class III
and Class IV Bond albedos are not significantly affected by condensates,
the columns referring to fractional condensation models are left blank.
The existence of Class V is a result of very strong
stellar insolation, so ``isolated'' T-P profile models are not calculated.  In all cases,
non-equilibrium condensates are ignored.}
\enddata
\end{deluxetable}

\clearpage

\begin{deluxetable}{cccccc}
\tablewidth{16cm}
\tablenum{1b}
\tablecaption{Estimated Bond Albedos of EGPs \label{Bondtabb}}
\tablehead{
\colhead{star} & \colhead{EGP class} & \colhead{fiducial}
& \colhead{``isolated''} & \colhead{10\% cond.}
& \colhead{1\% cond.}}
\startdata

G7V & I   & 0.55 & 0.58 & 0.52 & 0.44 \nl
    & II  & 0.79 & 0.79 & 0.69 & 0.34 \nl
    & III & 0.10 & 0.07 &     &     \nl
    & IV  & 0.02 & 0.15 &     &     \nl
    & V   & 0.55 &      &     &     \nl
K4V & I   & 0.48 & 0.52 & 0.44 & 0.33 \nl
    & II  & 0.70 & 0.70 & 0.60 & 0.25 \nl
    & III & 0.05 & 0.04 &     &     \nl
    & IV  & $<0.01$ & 0.11 &     &     \nl
    & V   & 0.53 &      &      &      \nl
M4V & I   & 0.38 & 0.43 & 0.33 & 0.16 \nl
    & II  & 0.56 & 0.55 & 0.47 & 0.16 \nl
    & III & 0.01 & $<0.01$ &     &     \nl
    & IV  & $<0.01$ & 0.08 &     &     \nl
    & V   & 0.51 &      &      &      \nl

\enddata
\end{deluxetable}

\clearpage

\begin{deluxetable}{ccccccc}
\tablewidth{16cm}
\tablenum{2a}
\tablecaption{Class II EGPs (``Water Cloud'') \label{EGPtab2a}}
\tablehead{
\colhead{object} & \colhead{star} & \colhead{M$_{\textrm{p}}$sin$i$ (M$_{\textrm{J}}$)}
& \colhead{{\it a\/} (AU)} & \colhead{cond.} & \colhead{{\it A$_B$}}
& \colhead{T$_{\textrm{eff}}$ (K)}}
\startdata

Gl 876b   &  M4V  &  1.9  &  $\sim$0.2   &  full  &  0.56  &  180 \nl
&&&& 10\%  &  0.47 &  182 \nl
&&&& 1\%   &  0.16 &  199 \nl
HR 5568b  &  K4V  &  0.75 &  $\sim$1.0   &  full  &  -     &  -   \nl
&&&& 10\%  &  -    &  -   \nl
&&&& 1\%   &  0.25 &  160 \nl
HD 210277b  &  G7V  &  1.28 &  $\sim$1.15  &  full  &  0.79  &  177 \nl
&&&& 10\%  &  0.69 &  194 \nl
&&&& 1\%   &  0.34 &  232 \nl
HR 810b   &  G0V  &  2.0  &  $\sim$1.2   &  full  &  0.82  &  192 \nl
&&&& 10\%  &  0.72 &  213 \nl
&&&& 1\%   &  0.38 &  254 \nl

\tablenotetext{}{Class II EGPs and their central stars, masses, and orbital
distances, along with estimated Bond albedos and effective temperatures, assuming
various condensation fractions.  Non-equilibrium condensates are ignored.
Internal luminosities are estimated using the evolutionary models of Burrows
et al. (1997) and assuming an age of 5 Gyr, except for HD 210277 b (8 Gyr),
47 UMa b (7 Gyr), and $\upsilon$ And d (3 Gyr).
An
absence of albedo and T$_{\textrm{eff}}$ entries indicates that for the expected
EGP Bond albedo assuming the given condensation fraction, T$_{\textrm{eff}}$ is low
enough such that condensed ammonia, rather than water, should reside in the
upper troposphere.  Hence, this combination
of parameters should not result in a Class II EGP.}

\enddata
\end{deluxetable}

\clearpage

\begin{deluxetable}{ccccccc}
\tablewidth{16cm}
\tablenum{2b}
\tablecaption{Class II EGPs (``Water Cloud'') \label{EGPtab2b}}
\tablehead{
\colhead{object} & \colhead{star} & \colhead{M$_{\textrm{p}}$sin$i$ (M$_{\textrm{J}}$)}
& \colhead{{\it a\/} (AU)} & \colhead{cond.} & \colhead{{\it A$_B$}}
& \colhead{T$_{\textrm{eff}}$ (K)}}
\startdata

16 Cyg Bb  & G2.5V &  1.66 &  1.7         &  full  &  0.81  &  158 \nl
&&&& 10\%  &  0.71 &  170 \nl
&&&& 1\%   &  0.37 &  198 \nl
47 UMa b   &  G0V  &  2.4  &  2.1         &  full  &  0.82  &  160 \nl
&&&& 10\%  &  0.72 &  172 \nl
&&&& 1\%   &  0.38 &  199 \nl
$\upsilon$ And d & F7V & 4.61 & 2.50      &  full  &  0.84  &  228 \nl
&&&& 10\%  &  0.74 &  233 \nl
&&&& 1\%   &  0.40 &  247 \nl
Gl 614b   &  K0V  &  3.3   &  2.5       &  full  &  0.75  &  168 \nl
&&&& 10\%  &  0.65 &  170 \nl
&&&& 1\%   &  0.30 &  177 \nl
55 Cnc c   &  G8V  &  $\sim$5 &  3.8      &  full  &  0.78  &  198 \nl
&&&& 10\%  &  0.68 &  199 \nl
&&&& 1\%   &  0.33 &  201 \nl
 
\enddata
\end{deluxetable}

\clearpage

\begin{deluxetable}{cccccc}
\tablewidth{16cm}
\tablenum{3}
\tablecaption{Class III EGPs (``Clear'') \label{EGPtab3}}
\tablehead{
\colhead{object} & \colhead{star} & \colhead{M$_{\textrm{p}}$sin$i$ (M$_{\textrm{J}}$)}
& \colhead{{\it a\/} (AU)} & \colhead{{\it A$_B$}}
& \colhead{T$_{\textrm{eff}}$ (K)}}
\startdata

HD 130322b  &  K0V  &  1.08  &   0.08    &  0.07  &  810 \nl
55 Cnc b   &  G8V  &  0.84  &   0.11    &  0.10  &  690 \nl
Gl 86 Ab   &  K1V  &  4.9   &   0.11    &  0.07  &  660 \nl
HD 195019b &  G3V  &  3.4   &   0.14    &  0.12  &  720 \nl
HD 199263b &  K2V  &  0.76  &   0.15    &  0.07  &  540 \nl
$\rho$ Cr Bb& G0V  &  1.13  &   0.23    &  0.13  &  670 \nl
HR 7875b  &  F8V  &  0.69  & $\sim$0.25&  0.14  &  650 \nl
HD 168443b &  G8IV &  5.04  &   0.277   &  0.10  &  620 \nl
HD 114762b &  F9V  &  $\sim$10    &   0.38    &  0.13  &  510 \nl
70 Vir b   &  G4V  &  6.9   &   0.45    &  0.11  &  380 \nl
$\upsilon$ And c&F7V& 2.11  &   0.83    &  0.14  &  370 \nl

\tablenotetext{}{Class III EGPs and their central stars, masses, and orbital
distances, along with estimated Bond albedos and effective temperatures.
Non-equilibrium condensates are ignored.}

\enddata
\end{deluxetable}

\clearpage

\begin{deluxetable}{cccccc}
\tablewidth{16cm}
\tablenum{4}
\tablecaption{Class IV EGPs (Roasters) \label{EGPtab4}}
\tablehead{
\colhead{object} & \colhead{star} & \colhead{M$_{\textrm{p}}$sin$i$ (M$_{\textrm{J}}$)}
& \colhead{{\it a\/} (AU)} & \colhead{{\it A$_B$}}
& \colhead{T$_{\textrm{eff}}$ (K)}}
\startdata

HD 187123b &  G3V  &  0.52  &   0.0415 &  0.03  &  1460 \nl
51 Peg b   &  G2.5V&  0.45  &   0.05   &  0.03  &  1240 \nl
$\upsilon$ And b&F7V& 0.71  &   0.059  &  0.03  &  1430 \nl
HD 217107b&  G7V  &  1.28  &   0.07   &  0.02  &  1030 \nl

\tablenotetext{}{Class IV EGPs and their central stars, masses, and orbital
distances, along with estimated Bond albedos and effective temperatures.
Non-equilibrium condensates are ignored.}

\enddata
\end{deluxetable}

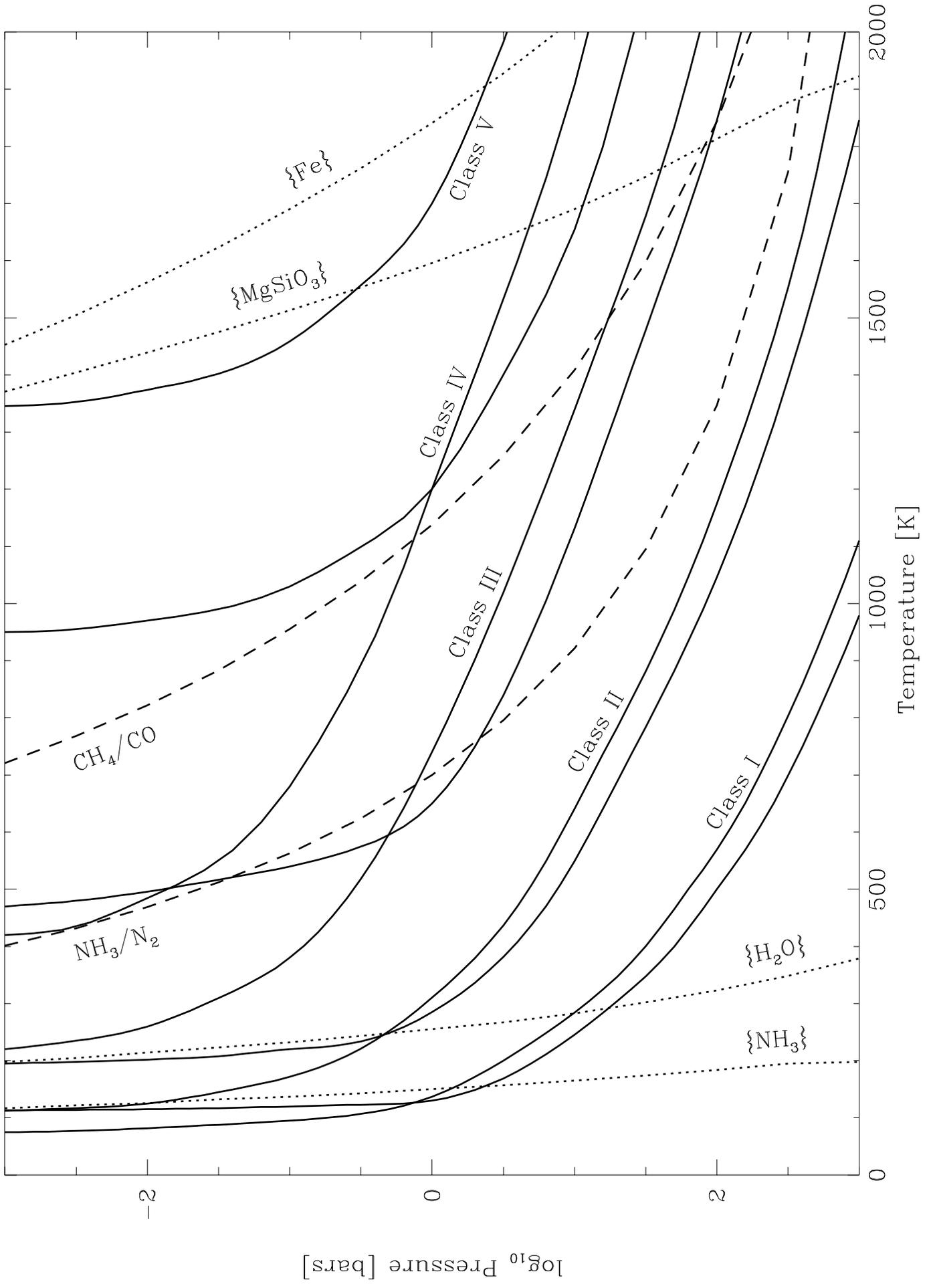
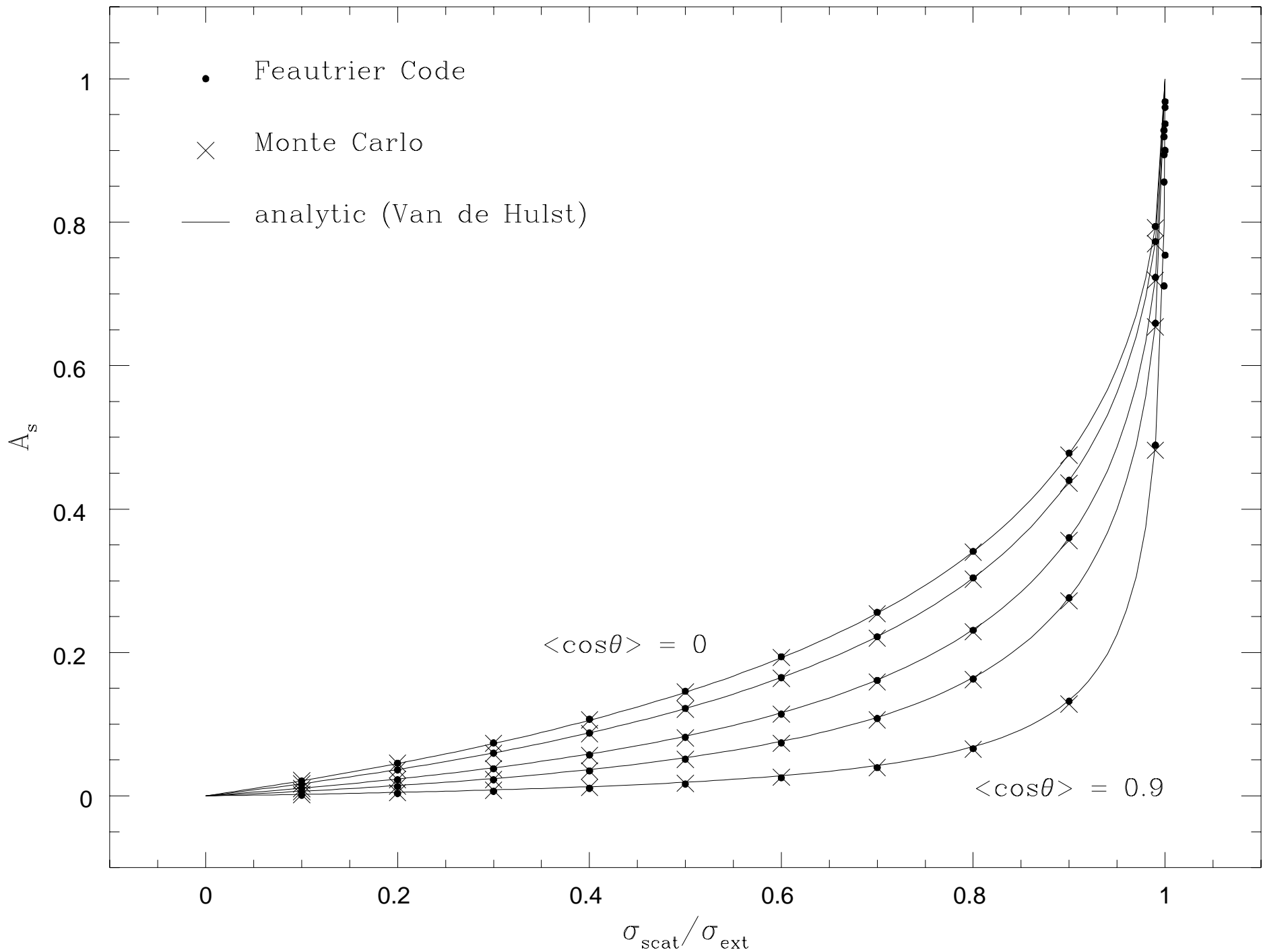
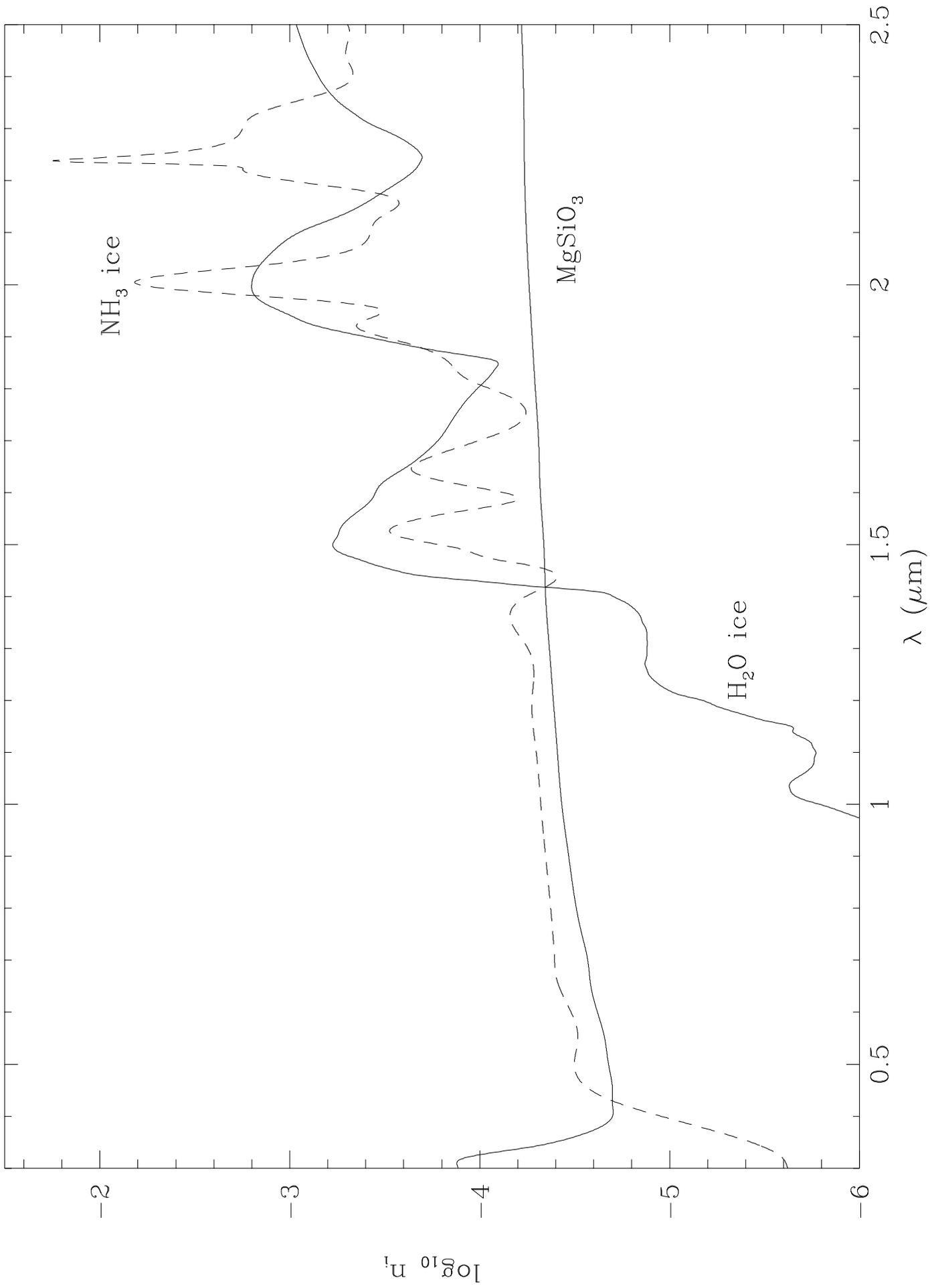
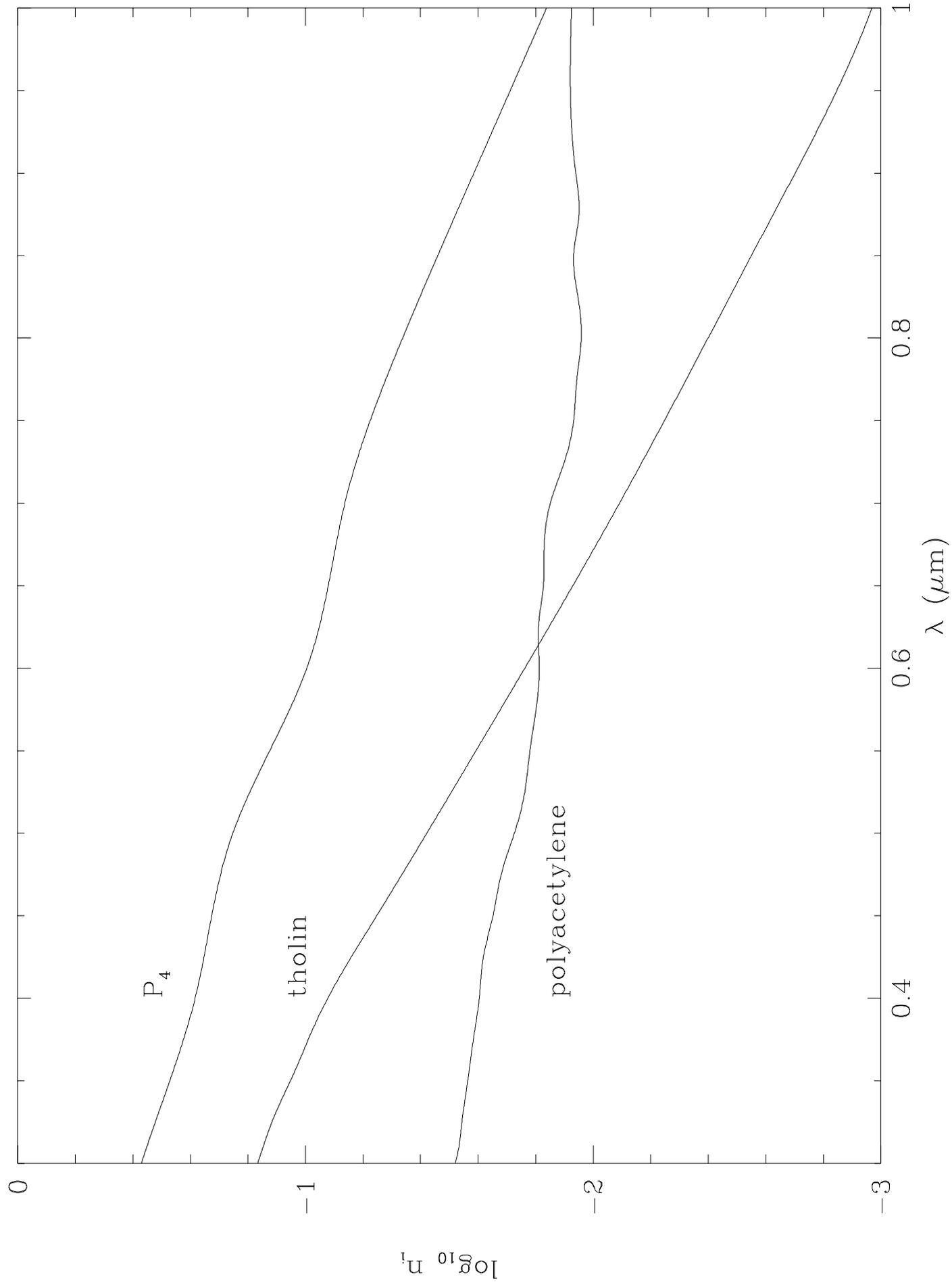
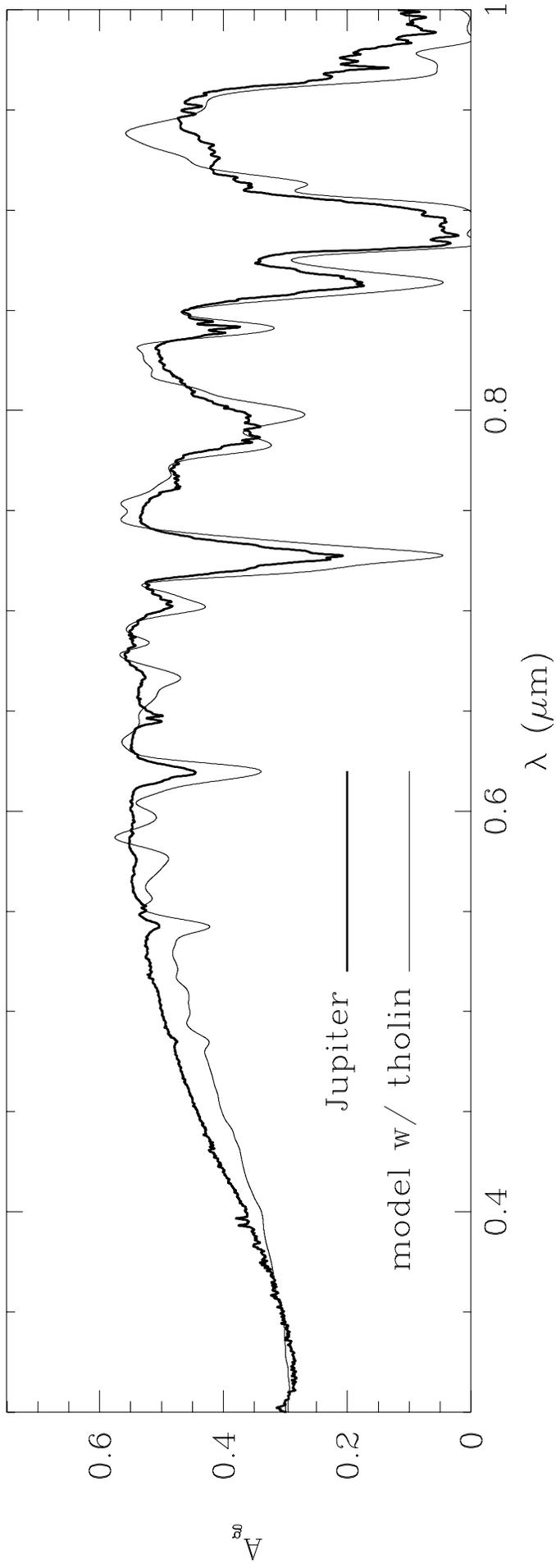
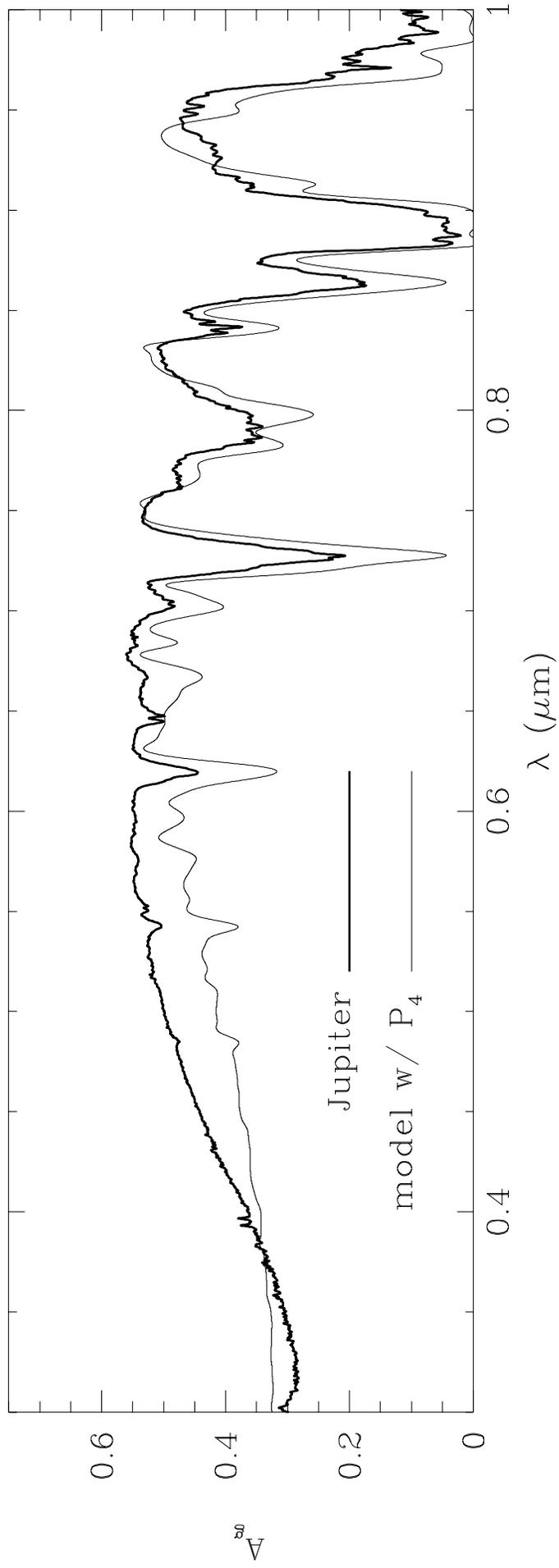
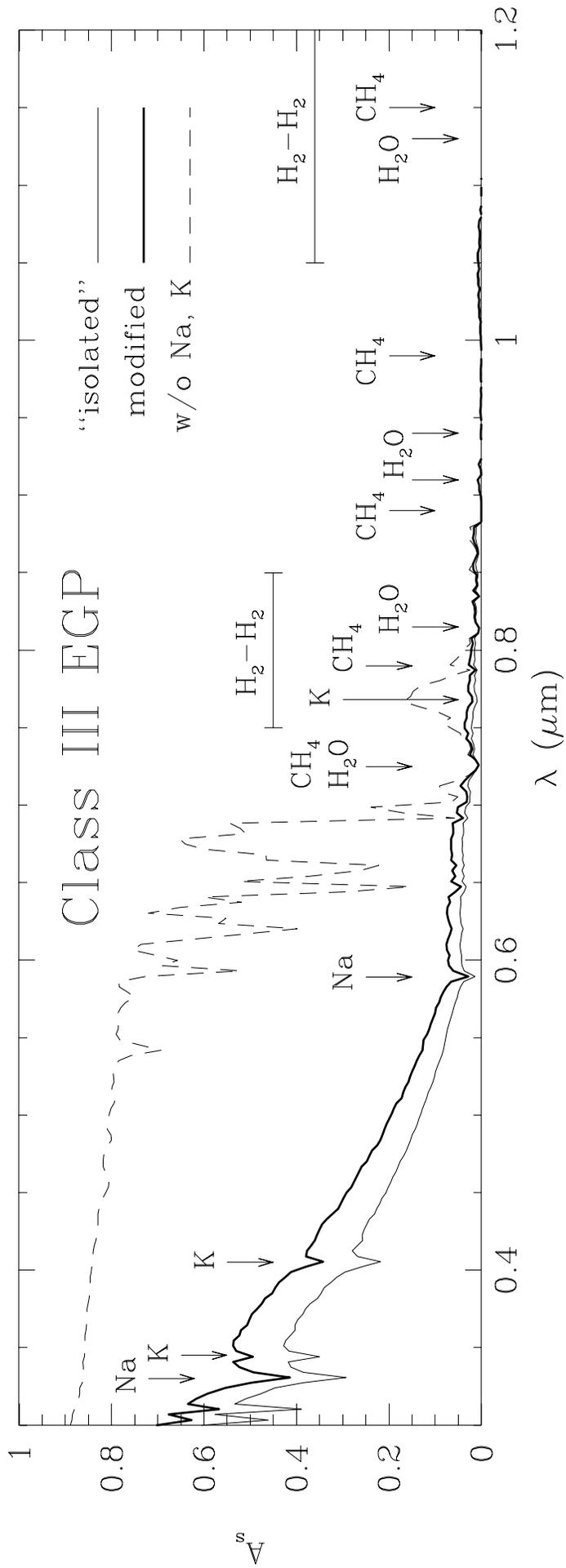
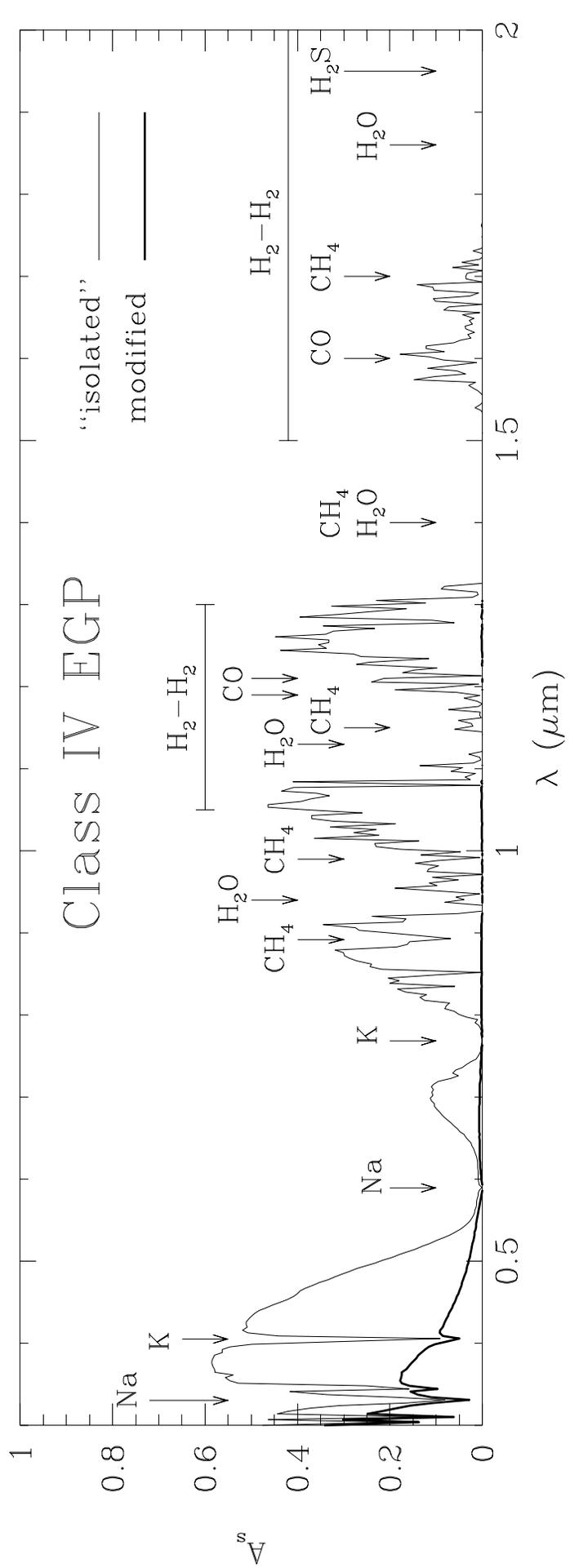
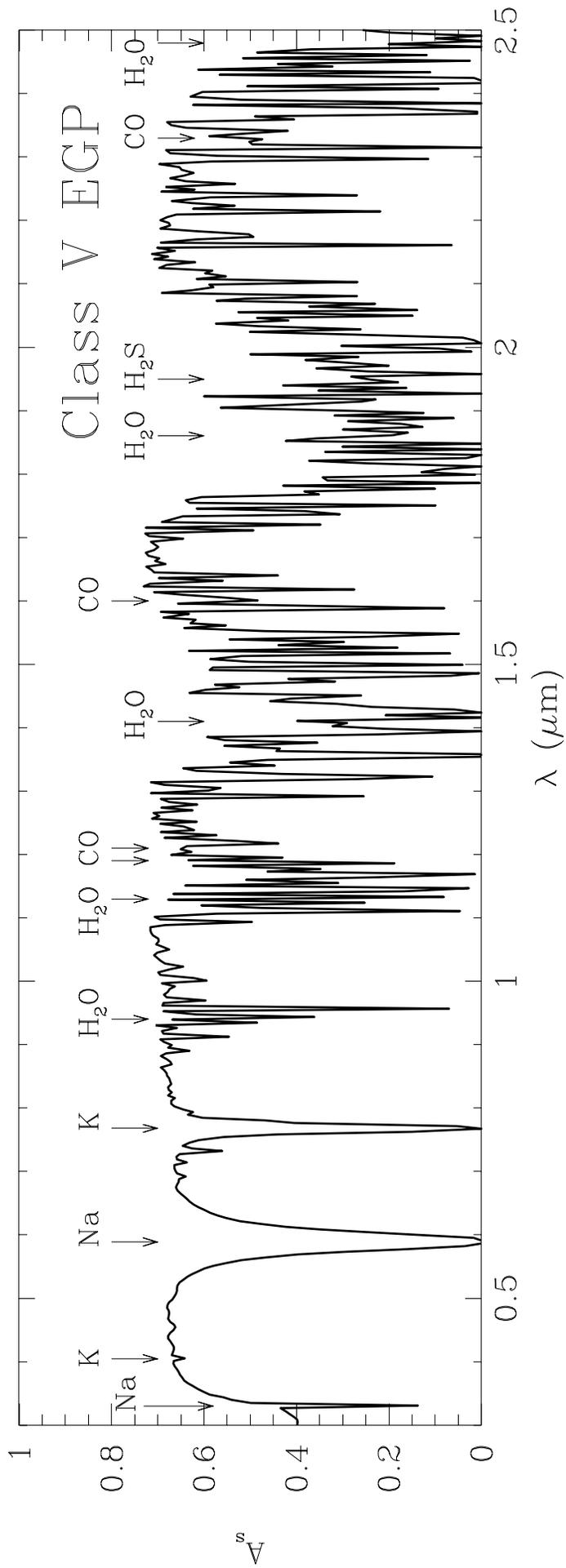
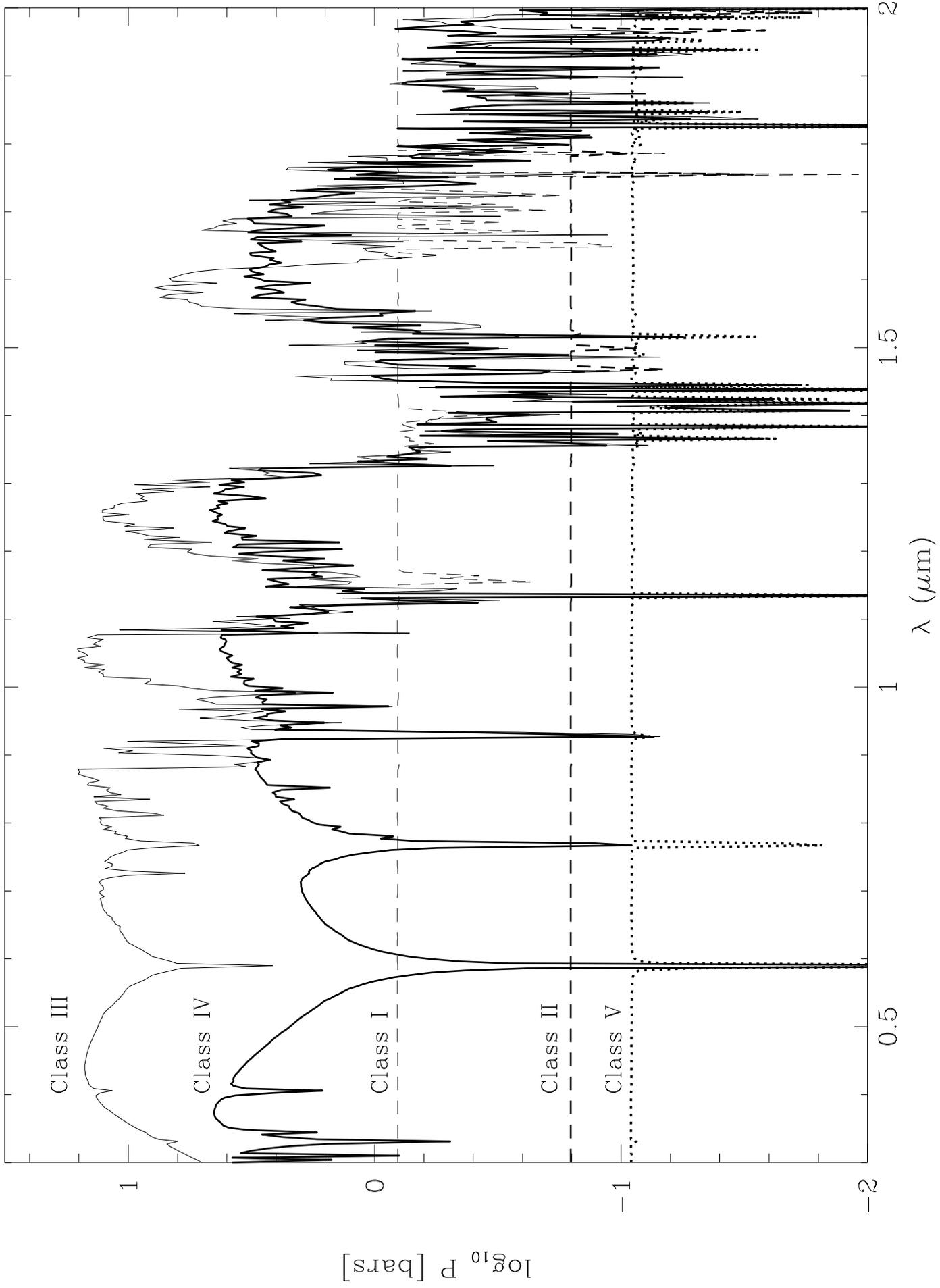
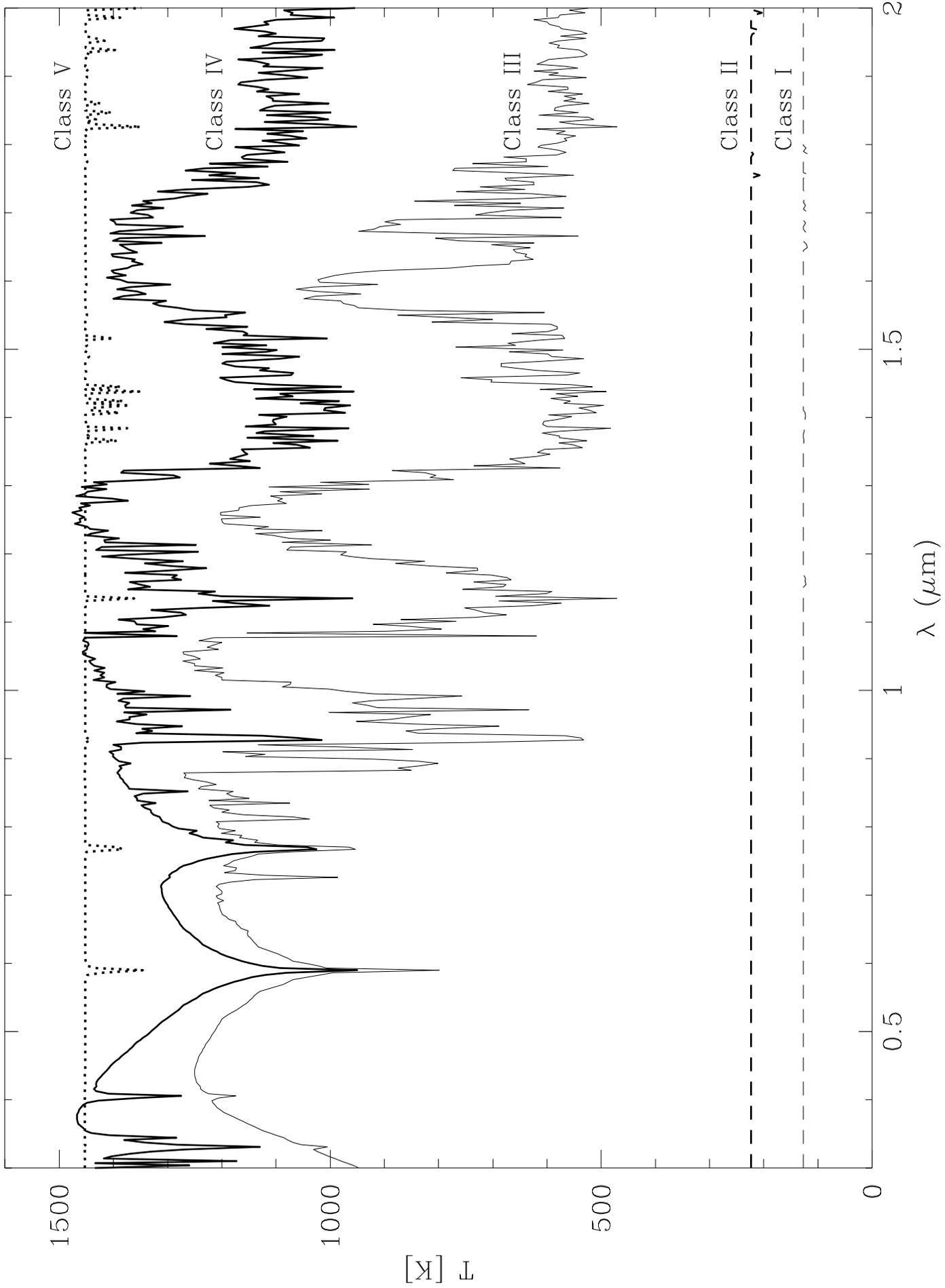
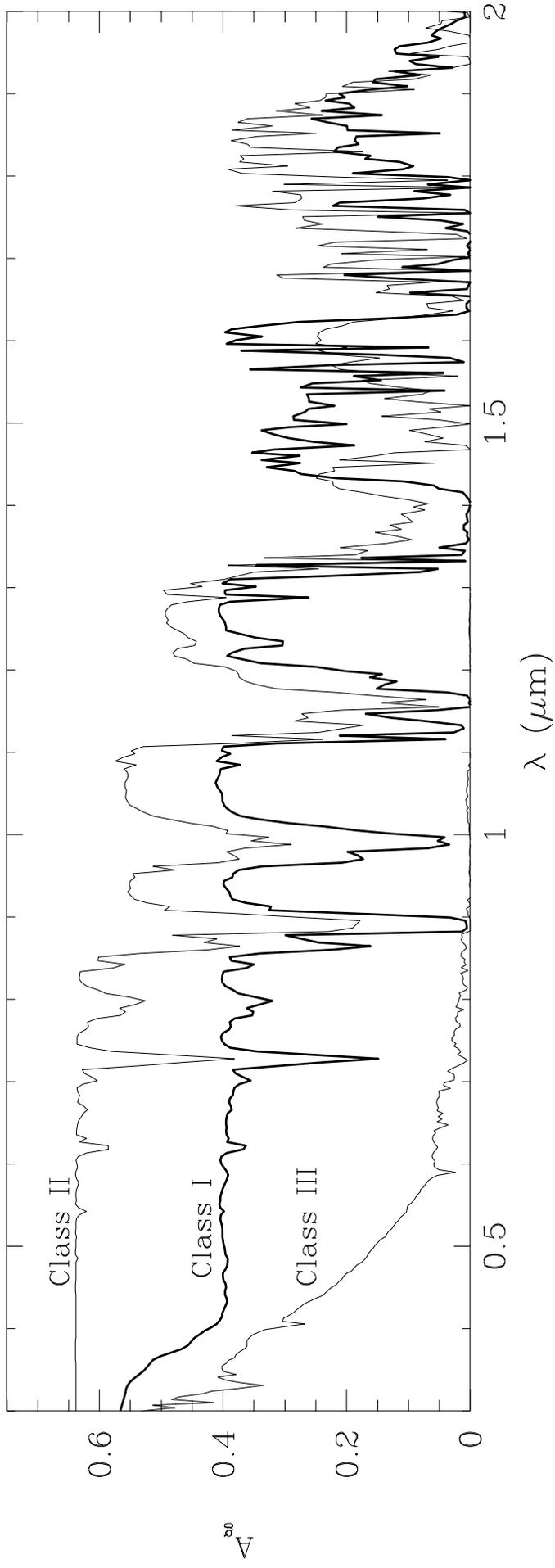
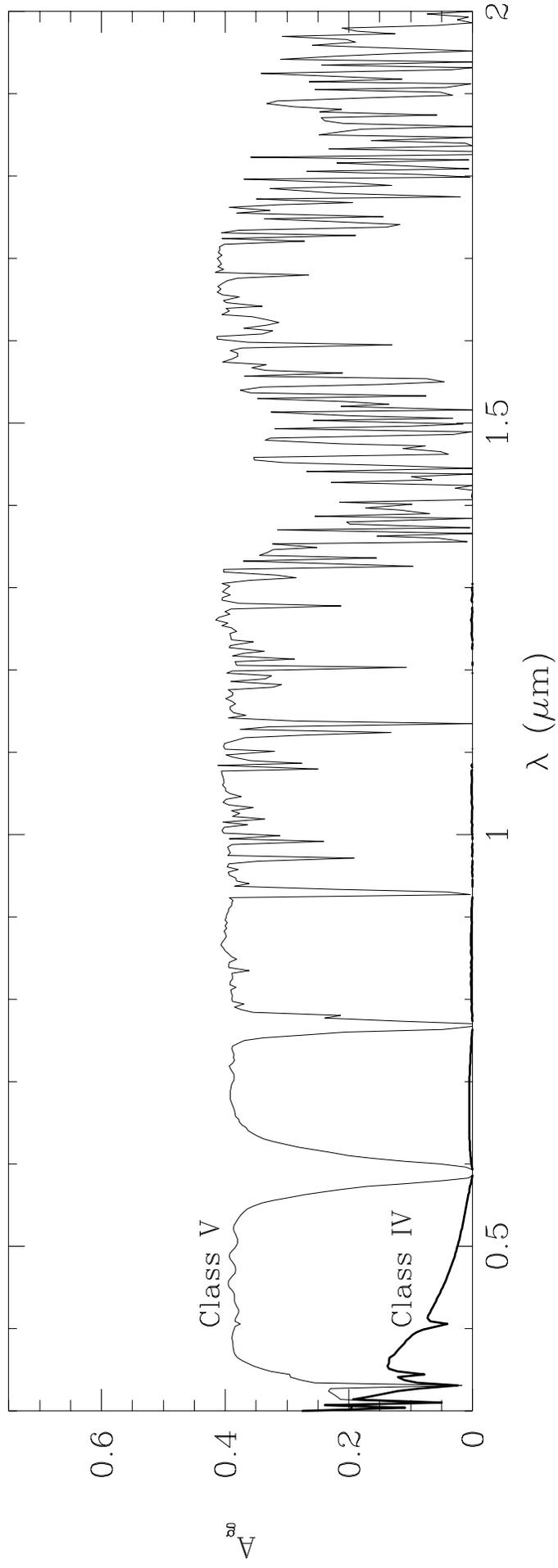
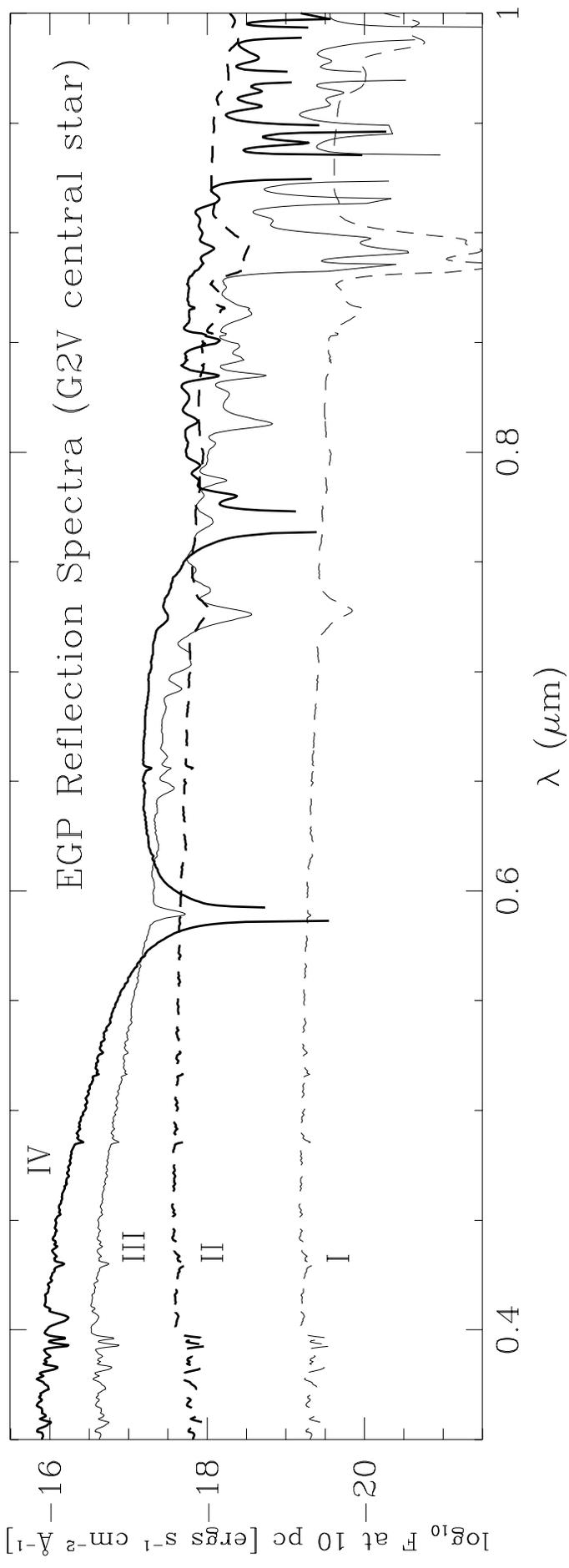
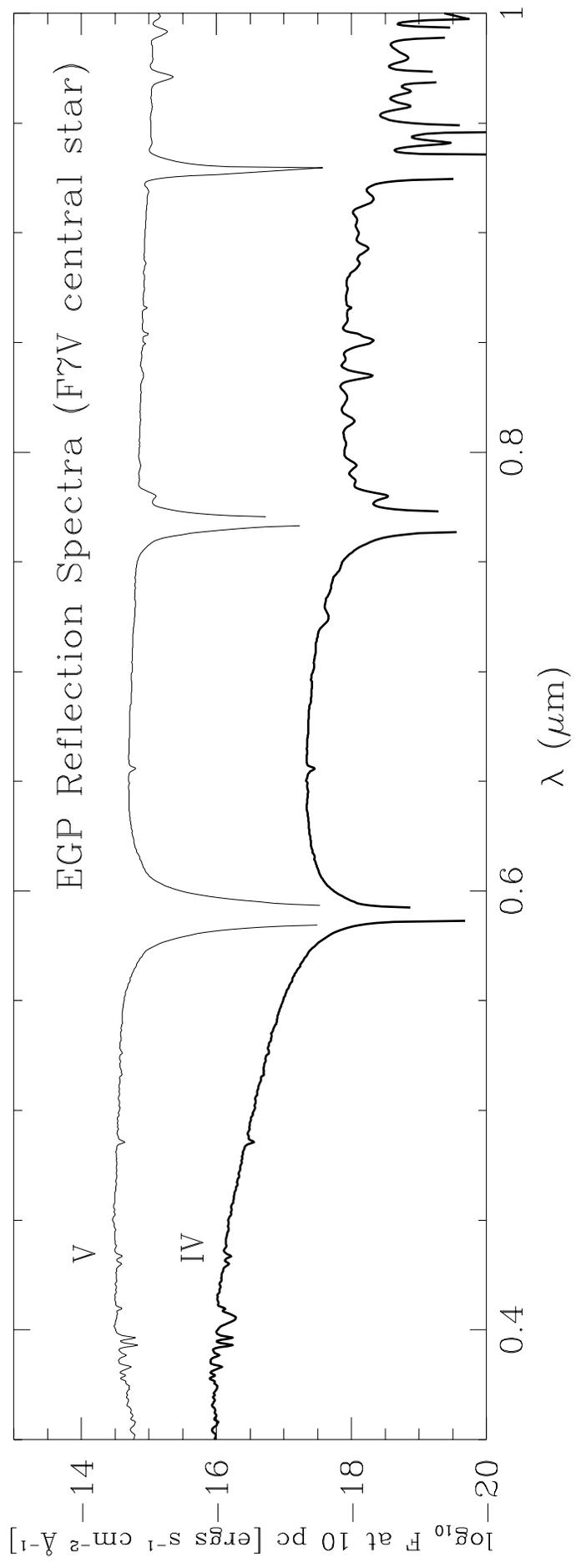
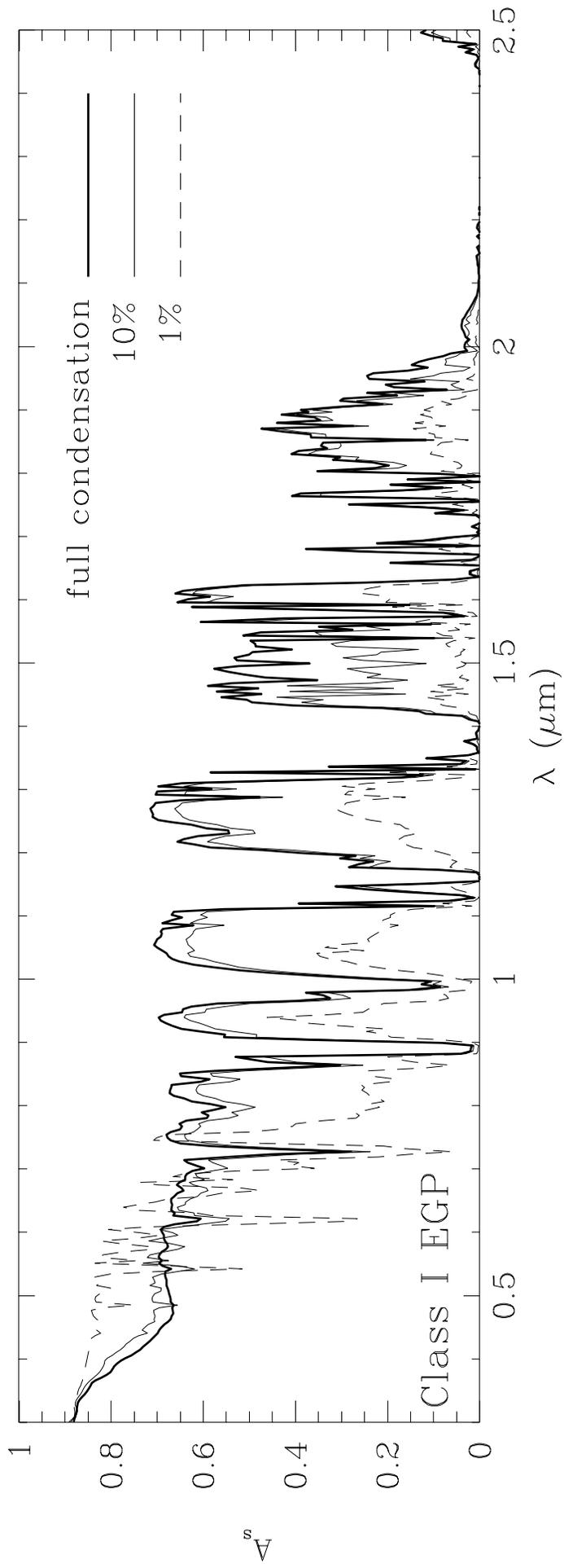
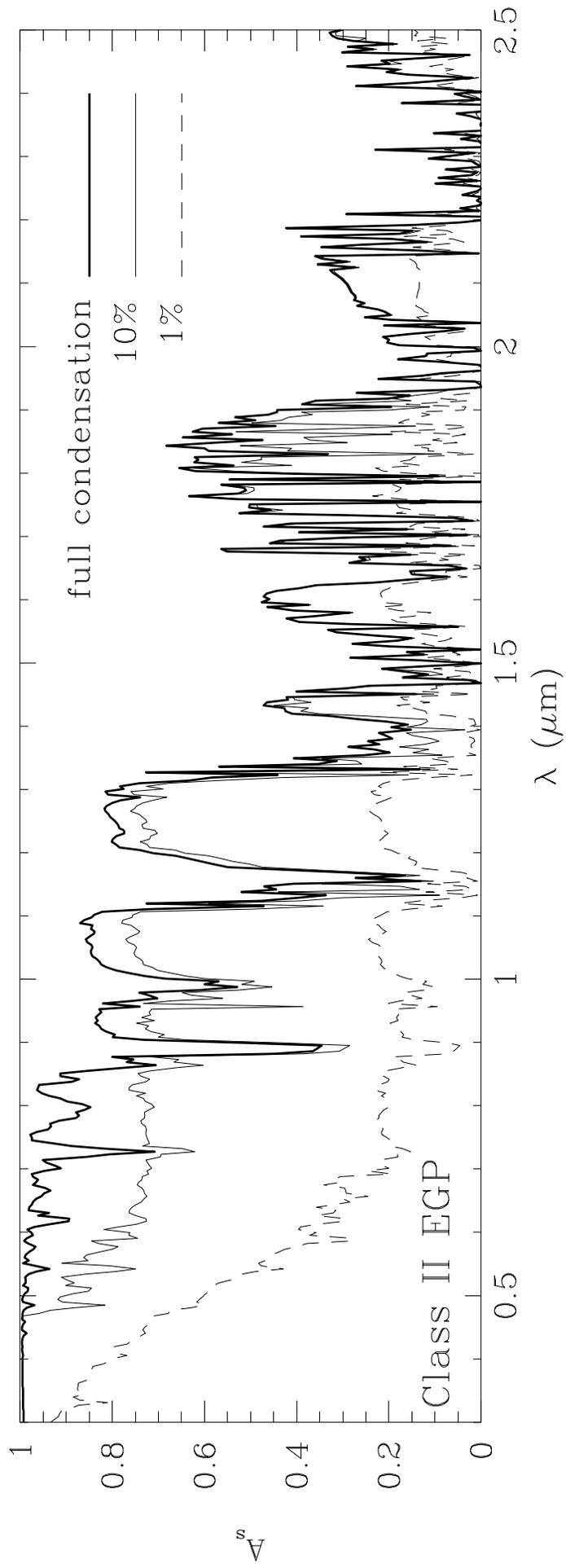
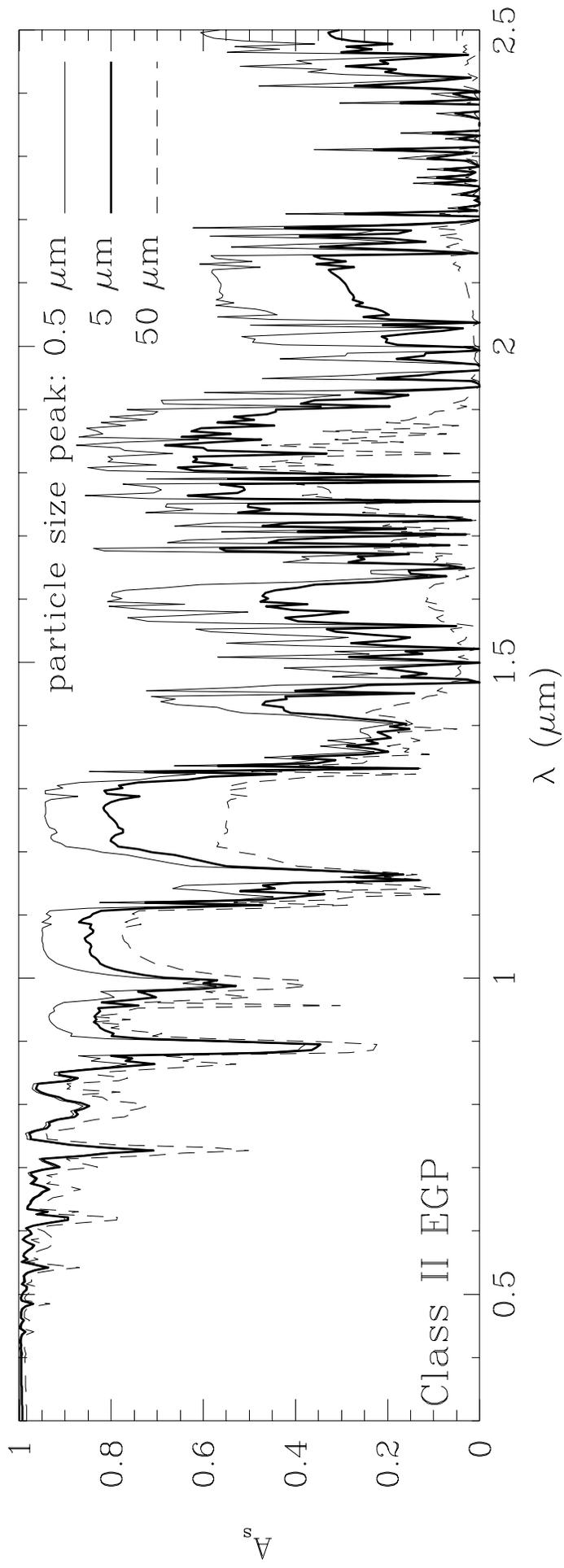
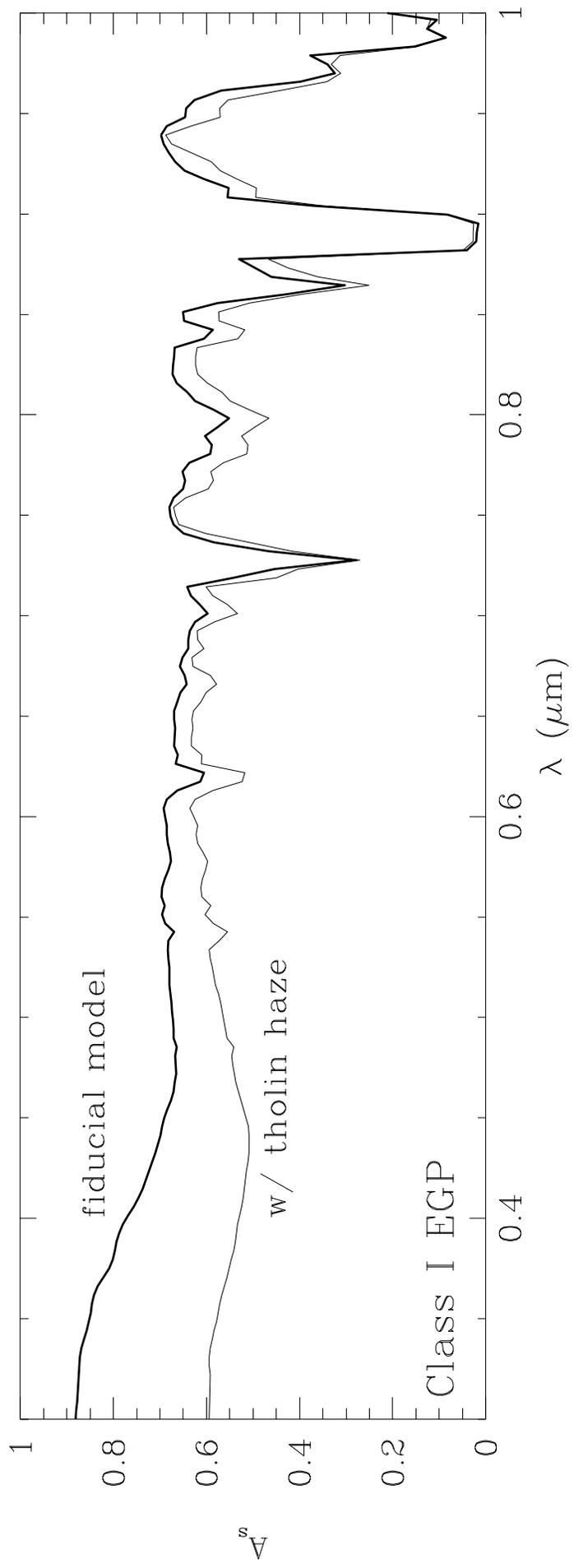
\end{document}